\documentclass[a4paper,twocolumn,11pt,accepted=2025-04-24]{quantumarticle}
\pdfoutput=1
\usepackage[utf8]{inputenc}
\usepackage[english]{babel}
\usepackage[T1]{fontenc}
\usepackage{amsmath}
\usepackage{hyperref}

\usepackage{tikz}
\usepackage{lipsum}

\usepackage[numbers,sort&compress]{natbib}
\usepackage[utf8]{inputenc}
\usepackage[english]{babel}
\usepackage[T1]{fontenc}
\usepackage{amssymb}
\usepackage{hyperref}
\usepackage{amsfonts}

\usepackage{graphicx}
\usepackage{dcolumn}
\usepackage{bm}
\usepackage{color}
\usepackage{blindtext}
\usepackage{titlesec}

\usepackage{simpler-wick}
\usepackage[shortlabels]{enumitem}
 \usepackage{capt-of}

\newcommand{\beqar}{\begin{eqnarray}}
\newcommand{\eeqar}{\end{eqnarray}}

\newcommand{\bcen}{\begin{center}}
\newcommand{\ecen}{\end{center}}

\newcommand{\lam}{\lambda}
\newcommand{\bra}[1]{\left< #1 \right|}
\newcommand{\ket}[1]{\left| #1 \right>}

\newcommand{\f}[2]{\frac{#1}{#2}}
\renewcommand{\b}[1]{\left({#1}\right)}

\renewcommand{\sb}[1]{\left[{#1}\right]}
\newcommand{\mean}[1]{\langle {#1} \rangle}

\newcommand{\ra}{\rightarrow}

\begin{document}

\title{Interplay between external driving, dissipation and collective effects in the Markovian and non-Markovian regimes}

\author{Roie Dann}
\email{roie.dann@mpq.mpg.de}
\affiliation{Max-Planck-Institut für Quantenoptik, Hans-Kopfermann-Str. 1, 85748 Garching, Germany}%
\orcid{0000-0002-8883-790X}

\begin{abstract}
Understanding how external driving and dissipation jointly influence the dynamics of open quantum systems is essential for advancing the study of non-equilibrium quantum phenomena and developing quantum technologies. The present study addresses the issue by exploring the behavior of open systems in driven optical setups coupled to a bosonic field. Starting from an exact non-Markovian master equation for linear systems, we extend the analysis to an ensemble of quantum emitters and validate the proposed solution. The analytical results unveil a range of intriguing phenomena, including pronounced non-Markovian corrections to the coherent driving and a collective cross-driving effect. These effects are experimentally accessible in platforms such as cavity QED, photonic crystals, and state-dependent optical lattices. In the Markovian limit, comparison with exact solutions reveal short-time non-Markovian effects that endure well beyond the environmental correlation decay time, alongside memory effects triggered by short laser pulses. These findings offer valuable insights into the dynamics of driven open systems, laying the groundwork for precise quantum state control.

\end{abstract}
\maketitle

\section{Introduction}

Driven open quantum systems are ubiquitous in quantum science \cite{breuer2002theory,weiss2012quantum,carmichael1999statistical,rivas2012open} and modern technological applications. Examples vary from free electrons \cite{gorlach2023high} and single atoms \cite{mabuchi2002cavity,walther2006cavity,hamsen2017two} to defects in bulk materials \cite{schirhagl2014nitrogen,liu2019coherent} and complex many-body systems \cite{bloch2008many,diehl2008quantum,zeiher2016many,sieberer2016keldysh,bluvstein2021controlling}. Applications span diverse fields, including quantum computing \cite{bluvstein2021controlling}, and sensing \cite{degen2017quantum,sekatski2016dynamical}. The dynamics of these systems are shaped by externally applied fields, often referred to as ``drive'' or ``control'', alongside unavoidable interactions with their surrounding environment. The intricate evolution and a wide array of physical phenomena are all encapsulated in the composite Hamiltonian

\begin{equation}
    H\b t= H_{S0}+H_d\b t + H_I +H_E~~.
\label{eq:H}
\end{equation}
where $H_{S0}$  represents the primary system's Hamiltonian, $H_{d}\b t$ 
 denotes the drive term, $H_I$, characterizes the system-environment interaction, and $H_E$  corresponds to the environment's Hamiltonian.

The structure of Eq. \eqref{eq:H} hints at the diverse range of phenomena that can emerge in driven open quantum systems. The non-commuting nature of the various Hamiltonian terms implies an inherent mutual influence among the system, drive, environment. For instance, the interplay among the primary system and environment gives rise to coherent dynamical effects, such as the well-known Lamb shift \cite{lamb1947fine,hansch1975doppler,rohlsberger2010collective,meir2014cooperative} and dispersive forces \cite{dung2002resonant,buhmann2013dispersion,gonzalez2015subwavelength}. Additionally, enduring environmental correlations can alter the reduced system dynamics through information backflow, leading to so-called  ``non-Markovian'' behavior \cite{rivas2014quantum,breuer2016colloquium,de2017dynamics}. These effects have been extensively studied, leading to proposed applications in quantum simulation \cite{chang2018colloquium,hung2016quantum}, laser physics \cite{bohnet2012steady}, and quantum networks \cite{goban2014atom}. However, the interplay between coherent driving, environmental coupling, and memory effects remain poorly understood.
Such interplay can potentially produce environment-dependent corrections to the drive as well as drive-dependent dissipation. The current gap in understanding is primarily due to the inherent complexity of the dynamics.

\begin{figure}[htb!]
\centering
\includegraphics[width=0.5\textwidth]{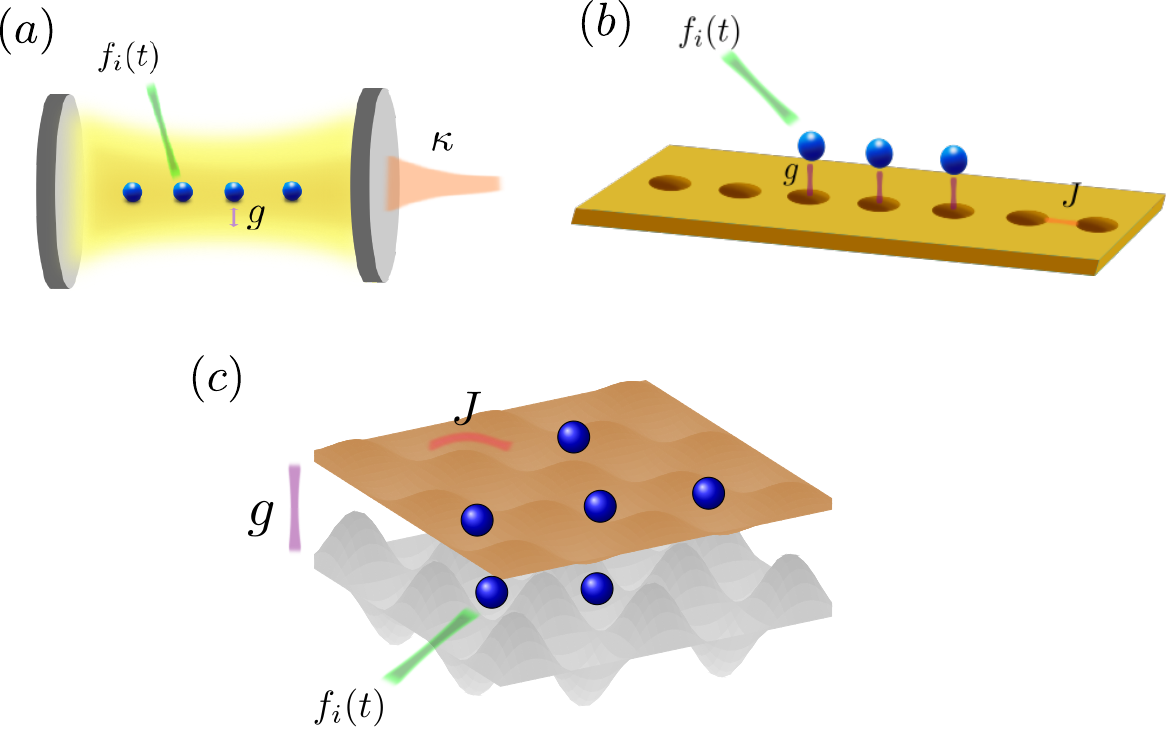}\\
\caption{There prominent of experimental setups allowing to witness the non-Markovian drive-dependent phenomena: (a) Cavity QED (b) emitters trapped near photonic crystals and (c) state dependent optical lattices.} 
\label{fig:exp_scheme} 
\end{figure}

The many-body nature of these composite systems along with the multitude of timescales involved restricts the number accurate solutions and limits perturbative analysis to specific physical regimes. In addition, the presence of the coherent drive introduces further theoretical challenges in evaluating the reduced dynamics of the primary system. The explicit time dependence of the drive breaks time-translation symmetry, formally necessitating a time-ordering procedure to obtain an accurate treatment.

The present contribution addresses these challenges by employing an analytical non-Markovian master equation to analyze the dynamics of both linear and nonlinear driven open optical systems. Specifically, we examine: (i) a system of bosonic modes coupled to a one-dimensional photonic crystal \cite{goban2014atom,lodahl2015interfacing,gonzalez2017markovian}, with dynamics governed by a quadratic composite Hamiltonian; and (ii) a collection of two-level emitters coupled to a bosonic environment. Our focus is on drive-dependent phenomena, emphasizing the intricate interplay between coherent driving, dissipation, and memory effects.

We begin by examining drive-dependent non-Markovian effects in linear systems, where a complete analytical solution can be derived (Sec. \ref{sec:linear_sys}). Building on the linear solution, we then construct an equation of motion for a system of quantum emitters (a non-lineary system) (Sec. \ref{sec:lin_ME}), which we term the {\textit{Linear Master Equation}} (LME). We emphasis that of the term ``linear'' refers to the fact that derivation of the master equation  is based on the dynamical analysis of linear system.

The derivation of the LME relies on the assumption that the environment’s influence on the primary system is only minimally impacted by nonlinear effects. A priori, this approach is expected to hold in the low-excitation regime, where emitter nonlinearities are relatively insignificant \cite{porras2008collective}. However, comparison with an exact pseudo-mode solution \cite{garraway1997decay,tamascelli2018nonperturbative} reveals that the LME remains accurate even under moderate driving conditions, where the transition of the two-level system saturates (Sec. \ref{subsec:strong_driving_single_emiter}). Additionally, the LME serves as a robust approximation in cases of strong system-environment coupling, where memory effects are prominent (Sec. \ref{subsec:non_markov_regime}
).

Crucially, the simple form of the LME enables a clear distinction between the various dynamical processes. Notably, the interaction between coherent driving and memory effects results in a non-Markovian self-correction to the original driving term (Sec. \ref{subsec:single harmonic}). Furthermore, when multiple linear modes are coupled to the same bosonic non-Markovian environment, the interaction generates cross-driving effects, where a laser applied to one mode effectively drives others (Sec. \ref{sec:two_modes}). Finally, the structure of the master equation predicts that the dissipation is independent of the drive.

These effects are readily observable in optical systems coupled to non-Markovian environments, including advanced experiments in cavity QED \cite{kimble1998strong,walther2006cavity}, quantum emitters interfaced with photonic crystals \cite{lopez2003materials,lodahl2015interfacing}, and two-state optical lattices \cite{de2008matter,navarrete2011simulating} (Fig. \ref{fig:exp_scheme} and Sec. \ref{sec:experimental}).


In the Markovian regime, discarding information backflow from the environment greatly simplifies the dynamical problem \cite{davies1974markovian,breuer2002theory,rivas2012open,cohen1998atom}. 
However, even under this approximation, the interaction between coherent and incoherent dynamical components remains only partially understood. Analyses across various driving regimes offer varying conclusions into how external driving impacts the dissipation
\cite{scully1999quantum,szczygielski2013markovian,albash2012quantum,dann2018time}.  Characterizing this effect is essential for the development of open system control techniques \cite{khodjasteh2010arbitrarily,koch2016controlling,dann2019shortcut,dann2020fast,kuperman2020mechanical,dupays2020superadiabatic,dupays2021shortcuts}, noise mitigation \cite{cai2023quantum}, and the engineering of effective coherent dynamics in the presence of external noise \cite{kallush2022controlling,aroch2024mitigating}.

To clarify the intricate relationship between dissipation and external driving, we compare the exact evolution to the Lindblad master equation (LME) and other commonly used Markovian  treatments (Sec. \ref{sec:markov_regime}). These include the Optical Bloch \cite{scully1999quantum}, Adiabatic \cite{albash2012quantum}, Floquet \cite{ho1986floquet,kohler1997floquet,breuer1997dissipative,chu2004beyond,szczygielski2013markovian,szczygielski2014application,elouard2020thermodynamics} and Time-Dependent master equations \cite{di2024time}.

Our findings indicate that even under strong driving, dissipation remains unaffected by the drive and is best represented in terms of transition between the bare system eigenstates rather than a dressed basis. 
However, novel drive-dependent non-Markovian effects emerge for sufficiently short laser pulses, where the bandwidths of the environment and the drive become comparable (Sec. \ref{subsubsection:bandwidth}). 
Additionally, comparison with the exact solution reveals that non-Markovian effects can manifest at times significantly longer than the memory decay time (Sec. \ref{subsec:strong_driving_single_emiter}). 

The manuscript concludes with an analysis of three experimental setups that may showcase drive-dependent non-Markovian phenomena (Sec. \ref{sec:experimental}). This is followed by a discussion of the key findings, potential extensions, and future directions for research.

\section{Background}

Under typical circumstances, the reduced dynamics of the open system are deduced by relying on a weak coupling with the environment and rapid decay of environmental correlations \cite{breuer2002theory,cohen1998atom},  leading to a Markovian description.
Nevertheless, with the introduction of structured environments, such as photonic crystals \cite{john1987strong,john1994spontaneous,goban2014atom,joannopoulos1997photonic} and nanophotonic structures  \cite{douglas2015quantum,chang2018colloquium}, moderate system-environment coupling \cite{prior2010efficient,matsuzaki2011magnetic,chin2012quantum,huelga2013vibrations,esposito2015quantum,roccati2024controlling}, and increasingly short laser pulses, a theoretical treatment beyond the Markovian regime is required \cite{rivas2014quantum,breuer2016colloquium,de2017dynamics}. 
A comprehensive dynamical description is essential for both the analysis of natural phenomena in biological light harvesting complexes \cite{blankenship2011comparing,huelga2013vibrations}, and solid state systems \cite{holstein1959studies,hubbard1964electron}, as well as in the development of quantum control tools \cite{rebentrost2009optimal,goban2014atom,koch2016controlling}, single-photon sources \cite{arcari2014near,hood2016atom}, and sensing techniques \cite{chin2012quantum,matsuzaki2011magnetic}.

\subsubsection{Non-Markovian 
Dynamics}

Non-Markovian dynamics of driven non-linear open systems have been studied extensively using a variety of numerical techniques. 
Quasi-adiabatic \cite{makri1995tensor,sahrapour2013tunneling} and non-interacting/inter-blip path integral methods \cite{leggett1987dynamics,grifoni1995cooperative,winterstetter1997dynamical}, hierarchical equations of motion \cite{meier1999non,jin2008exact,wang2013exact,xu2023performance}, generalized master equations \cite{grifoni1996exact,magazzu2018asymptotic}, renormalization group methods \cite{shapourian2016dynamical}, driven Liouville von-Neumann formalism \cite{hod2016driven} and  stochastic  Schr\"odinger equation \cite{orth2013nonperturbative} have been employed to study the dynamics of a variety of systems. Specifically, a large effort has been devoted to the analysis of the spin-boson model, which serves as a toy model for chemical processes involving two isolated quantum states coupled to a complex environment such as a solvent or molecular vibrations \cite{leggett1987dynamics,weiss2012quantum}.

The numerical methods are often both precise and efficient in a defined physical regime, such as moderate temperature or up to second order in the system-environment coupling strength. Moreover, they compute system expectation values. Importantly,  even when the dynamics of the expectation values can be evaluated accurately, it remains extremely difficult to differentiate and analyze the underlying dominant physical processes. This information is readily obtained by analyzing the form of an analytical equation of motion, where the role of different terms can be traced to distinct physical process. 

Notably, Ref. \cite{amati2024dynamical} employed a numerical solution of the hierarchical equations of motion to investigate signatures of non-Markovianity in the periodically driven spin-boson model.  These include negativity in the decay rates of a time-local master equation and memory kernel decay time of a generalized Lindblad master equation.
Alternatively, previously proposed analytical approaches for non-linear systems have relied on either the slow variation of the spectral density \cite{cao2021non}, or a high driving frequencies \cite{grifoni1996exact,hartmann2000driven}.

The dynamics of non-driven open linear systems have been studied extensively \cite{caldeira1983quantum,unruh1989reduction,hu1992quantum,lambropoulos2000fundamental,xiong2010exact,wu2010non,zhang2012general}, while driven systems have been studied in the context of bosonic \cite{lei2012quantum} and fermionic \cite{tu2008non} transport models, coherent control \cite{quang1997coherent} and the derivation of an exact master equation \cite{yang2013master}.

\subsubsection{Markovian Dynamics}

Several constructions of Markovian master equations for driven open systems have been introduced. They typically focus on specific driving regimes and follow the Born-Markov-secular approximations \cite{breuer2002theory} or employ projection operator techniques \cite{nakajima1958quantum,zwanzig1960ensemble}. For slow driving, the approach leads to an adiabatic master equation \cite{albash2012quantum}, where the reduced dynamics follows the instantaneous composite Hamiltonian. The physical interpretation, in this case, involves environment induced transitions in an instantaneous eigenbasis of the primary system and drive. For rapid oscillatory driving, the Floquet theorem leads to the Floquet Master Equation (FME) \cite{ho1986floquet,kohler1997floquet,breuer1997dissipative,chu2004beyond,szczygielski2013markovian,szczygielski2014application}.
The derivation of the FME uses a Fourier decomposition of the system interaction operators in the interaction picture. Such decomposition leads to a Lindblad form with drive dependent kinetic coefficients \cite{elouard2020thermodynamics}. Here, the dissipation is captured by transitions between the Floquet states, which for a monochromatic drive correspond to the so-called dressed states \cite{cohen1998atom}.

Recently, generalizations for non-periodic and non-adiabatic drives have been proposed  
\cite{yamaguchi2017markovian,dann2018time,mozgunov2020completely,di2024time}. Notability, Ref. \cite{di2024time} combines the projection operators technique with a rescaling of time, in the spirit of the seminal work of Davies \cite{davies1974markovian}. This procedure leads to what is denoted here as the Time-Dependent  Master Equation (TDME). In the interaction picture, relative to the primary system dynamics and drive, the TDME is characterized by adiabatic Lindblad jump operators and kinetic coefficients. The non-adiabaticity in this approach arises when transforming back to the Schrödinger picture, performed using the numerically evaluated exact time-evolution operator for the isolated system. 

The different master equations may generally differ in both their operator structure and kinetic coefficients. As a result, they commonly provide distinct physical predictions, even within the same driving regime. This diversity stems different ways the basic physical approximations can be implemented. Moreover, change of the order of approximations may also modify the final result.

An axiomatic approach has been developed to resolve this ambiguity \cite{dann2021quantum,dann2022non}. Within this framework, a number of axioms are postulated and utilized to prove the structure of the corresponding master equation. However, the construction is based on  strong dynamical symmetries of the composite Hamiltonian, which in practice may not necessarily hold.

\section{Framework}
\label{sec:framework}

The validity of the fundamental Hamiltonian, Eq. \eqref{eq:H}, is illuminated by considering the quantum nature of the driving field. Specifically, the time-dependent representation of the drive ($H_d\b t$) can be derived from an autonomous (time-independent) description. This involves a transformation from an interaction picture where the Hamiltonian is time-independent and the field initially resides in a product of coherent states \cite{cohen1998atom}, {\footnote{For a detailed derivation see exercise 17 of \cite{cohen1998atom}}}. Consequently, the explicit time-dependent Hamiltonian does not relay on a semi-classical approximation and remains precise even for a weak coherent drive. Notably, this approach accounts for both the field’s quantum fluctuations and the back-action of the system on the field. The representation holds in the non-relativistic regime and for moderate field intensities satisfying $g_{SL}\ll \omega_S$, (see definitions below)
\footnote{Within the Coulomb gauge, the minimal coupling Hamiltonian includes a term proportional to the square of the electromagnetic vector potential $H_{I2}\sim q\boldsymbol{A}^2/m$ where $q$ and $m$ are the particle charge and mass. For moderate intensities, the small ratio between $H_{I2
}$ and the particle-field interaction $\f{H_{I2}}{H_{I1}}\sim\f{q^{2}\boldsymbol{A}^{2}/m}{qp\boldsymbol{A}/m}=\f{qp\boldsymbol{A}/m}{p^{2}/m}\sim\f{H_{I1}}{H_{P}}$, allows neglecting $H_{I2}$. Here, $p$ is the particle momentum and $H_P$ is the free particle Hamiltonian term.}.



The characteristic reduced primary system dynamics is determined by the relation between a restricted number of characteristic energy scales: (a) the typical frequency of the system $\omega_S$, (b) detuning with respect to the central laser frequency $\Delta = \omega_S - \omega_L$, (c) driving bandwidth $\Delta \omega_L$,  (d) environment spectral width $\Delta \omega_E$, (e) inverse of the local density of environment states $1/D\b{\omega_S}$, as well as (f) the system-environment coupling $g_{SE}$, and (g) the system-drive coupling $g_{SL}$. Our main focus is on system-environment Hamiltonian terms which conserve the number of excitations (U(1) symmetry). In this case a transformation to a rotating frame highlights that open system dynamics are independent of the magnitude of $\omega_S$ \footnote{For an Hamiltonian with a U(1) symmetry, it is convenient to work a frame rotating at the central laser frequency $\omega_L\approx \omega_S$. The rotation amount to shifting the spectrum of the system and environment and adding oscillating phases to the drive. Note, The validity conditions of the rotating wave approximation, leading to the U(1) symmetry, and the mapping of an autonomous Hamiltonian to an explicitly time-dependent one coincide.
}.
Additional factors which influence the dynamical behavior are the interplay between the light polarization and the spatial dimensions of the medium \cite{lodahl2017chiral} as well as retardation effects \cite{sinha2020non}. Since the present focus concerns the dynamical effects of the drive, these considerations are neglected.

The dynamics is commonly characterized by the significance of the memory effects.
Markovian evolution occurs when the environment's spectral features vary on a scale much larger than the system's  transition linewidth, i.e.,  
\begin{equation}
 g_{SE}^2 D\b{\omega_S}\ll \Delta \omega_E ~~.   
\label{eq:markov_cond}
\end{equation}
This leads to memoryless dynamics, where information backflow can be neglected, and the dissipation is characterized by an exponential decay with a rate given by Fermi's Golden Rule (FGR) $\Gamma_{\text{FGR}}\sim g_{SE}^2 D\b{\omega_S}$. Within this regime, the system-laser coupling is also  typically small with respect to the environment's band width, $g_{SL}\ll\Delta \omega_E$. Nevertheless, for a short pulses one can reach cases for which $\Delta \omega_L\sim \Delta \omega_E$, while  still maintaining the   Markovian condition (Eq. \eqref{eq:markov_cond}).

Beyond the Markovian regime, the restriction given by \eqref{eq:markov_cond} is lifted, and the system evolution may exhibit information backflow, resulting in oscillatory behavior \cite{rivas2014quantum}, power-law decay rates \cite{wodkiewicz1976markovian,knight1978non}, strong dispersive interactions \cite{gonzalez2017markovian} and fractional decay \cite{john1987strong,john1990quantum,yablonovitch1987inhibited,bykov1975spontaneous,kurizki1990two,shi2016bound,gonzalez2017markovian}. 
In the extreme case, where $g_{SE}>\Delta \omega_E$, the impact of information backflow suppresses dissipation.



Finally, the present work focuses on optical systems interacting with the electromagnetic field. For these systems, it is well justified to consider a bosonic environment at $T=0$. Nevertheless, all the results can be generalized to the case of $T>0$ in a straightforward manner; see the discussion section and Appendix \ref{apsec:ME_parameters}. In addition, a similar theoretical analyses applies for a wide range of physical systems, such as solid-state color centers interacting with lattice phonon modes \cite{doherty2013nitrogen,goldman2015phonon}, molecular electronic or spin states coupled to molecular or solvent vibrational modes \cite{pachon2011physical,ishizaki2012quantum} and many more \cite{lodahl2015interfacing,blais2021circuit,barreiro2011open,wang2024simulating,byczuk2008correlated,anders2010dynamical,strand2015nonequilibrium,scarlatella2021dynamical}.

\section{Linear optical systems}
\label{sec:linear_sys}
We first consider a linear system of $N_S$ coupled resonators with frequencies $\omega_1,\omega_2,\dots,\omega_{N_S} $ 
interacting with a bosonic field initially in the vacuum state. 
Physical realizations of the theoretical model include driven coupled micro-resonators \cite{hajjaj2020linear}, mechanical oscillators \cite{jost2009entangled,kotler2021direct} and nano-structured photonic materials \cite{lodahl2015interfacing}. 
The latter constitutes an especially attractive platform, where high controllability enables access to exotic dynamical regimes involving strong system-environment couplings, a high density of states and narrow environmental bandwidths. For further experimental considerations see Sec. \ref{sec:experimental}.



Within the rotating wave approximation, the composite linear system Hamiltonian takes the following form
\begin{multline}
    H\b t=  \sum_{ij} M_{ij}a^\dagger_ia_j+ \sum_i\b{ f_i\b t a_i^\dagger +f_i^*\b t a_i}\\\sum_{k,i}\b{ \eta_{ik} a^\dagger_i b_k +\eta_{ik}^* a_i b_k^\dagger}+ \sum_k \omega_k b_k^\dagger b_k~~,
    \label{eq:H_comp1}
\end{multline}
where  $a_i$ and $b_k$ are  system and environment annihilation operators, with $i,j=1,\dots,N_S$, and $k=1,\dots,N_E$. $k$ may account for both the mode's wavevector and polarization and $\hbar$ is taken to be unity throughout the paper. The first two terms represent the internal system couplings and drive, while the last terms correspond to the system-environment interaction and the free environment Hamiltonian, respectively. The following analysis considers bosonic modes; nevertheless, it applies as well to fermionic Hamiltonians of the form of  \eqref{eq:H_comp1} with  minor modifications, Appendix \ref{apsec:ME_parameters}.

Before pursuing the exact system dynamics, it is beneficial to establish the connection between the Hamiltonian parameters and the typical energy scales. The typical system energy scale, $\omega_S$, is associated with the characteristic magnitude of the eigenvalues of $\boldsymbol{M}$, while the environment's bandwidth, $\Delta \omega_E$, is given by the variance of $\omega_k$ in the continuum limit. $g_{SL}\sim |f_i\b t|$, while $\Delta \omega_L$ is defined as the standard deviation of the driving terms', $\{f_i\b t\}$,  Fourier transform, and $g_{SE}$ corresponds to the ratio of the spectral density and the density of states  $g_{SE}^2\sim \sum_k |\eta_{ik}|^2\delta\b{\omega-\omega_k}/D\b{\omega_S}$.

\subsection{Exact master equation}
\label{subsec:exact_lin_ME}

The master equation for a driven bosonic composite systems, whose evolution is governed by a quadratic Hamiltonian  was derived in \cite{zhang2012general} (detailed derivations and an extension to fermionic composite systems is given in \cite{yang2013master}). The construction employs a coherent state representation, which can be evaluated analytically for a quadratic Hamiltonian of the form \eqref{eq:H_comp1}. An exact Fokker-Planck equation for the dynamical propagator is then deduced and mapped to a time-convolution-less master equation. It is explicitly given by 
\begin{multline}
    \dot{\rho}_S\b t = - i\sb{\tilde{H}_S\b t,\rho_S\b t}\\ +\sum_{ij}\Gamma_{ij}^{\downarrow}\b t{\cal D}_{a_j,a_i^\dagger}\sb{\rho_S\b t} + \Gamma_{ij}^{\uparrow} \b t {\cal D}_{a_i^\dagger,a_j}\sb{\rho_S\b t}~~,
    \label{eq:dot_rho_s_lin}
\end{multline}
where ${\cal D}_{x,y}\sb{\bullet}\equiv x\bullet y-\{yx,\bullet\}/2$, with 
\begin{equation}
    \tilde{H}_S\b t = \sum_{ij} a^{\dagger}_i{\Omega}_{ij}\b t a_j+\sum_{i}\b{k_{i}\b t a_{i}^{\dagger}+k_i^*\b ta_i}~~.
\end{equation} 
Here the  relaxation rates $\Gamma_{ij}^{\downarrow}$ and $\Gamma_{ij}^{\uparrow}$, and coherent term $\Omega_{ij}$, given explicitly in Appendix \ref{apsec:ME_parameters}, 
 are determined by the time-dependent matrix elements  of the Keldysh non-equilibrium Green function $\boldsymbol{W}\b{t,0}$ \cite{schwinger1961brownian}. The Green function is an $N_S$ by $N_S$ matrix that satisfies the integro-differential equation 
\begin{equation}
    \dot{\boldsymbol{W}}\b{t,t'}+i\boldsymbol{M}\boldsymbol{W}\b{t,t'}+\int_ {t'}^t{d\tau \boldsymbol{K}\b{t-\tau}\boldsymbol{W}\b{\tau,t'}=0}~~,
    \label{eq:W_t}
\end{equation}
with initial condition $\boldsymbol{W}\b{ 0,0}=\boldsymbol{I}_S$, 
where $M_{ij}$ are elements of $\boldsymbol{M}$, $\boldsymbol{K}\b{t,t'}=\boldsymbol{K}\b{\tau=t-t'}=\int_0^{\infty}{d\omega \boldsymbol{J}\b{\omega}e^{-i\omega \tau}}$ is the memory kernel and the elements of the spectral density functions read $\sb{\boldsymbol{J}}_{jk}=\sum_{l}\eta_{jl}\eta_{kl}^{*}\delta\b{\omega-\omega_{l}}$. 
The drive terms $\{k_i\}$ are elements of the vector 
\begin{equation}
    \boldsymbol{k}\b t =\boldsymbol{f}\b t +\boldsymbol{f}_{NM}\b t~~,
    \label{eq:k_t}
\end{equation}
where the subscript $NM$ designates a Non-Markovian contribution
\begin{equation}
 \boldsymbol{f}_{NM}\b t=\int _0^t{\dot{\boldsymbol{W}}\b{t,\tau}\boldsymbol{f}\b{\tau}d\tau}-\dot{\boldsymbol{W}}\boldsymbol{W}^{-1}\tilde{\boldsymbol{f}}~~,  
 \label{eq:f_NM}
\end{equation}
with $\dot{x}\equiv dx/dt$,
$\tilde{\boldsymbol{f}}\b t=\int_0^t{\boldsymbol{W}\b{t,\tau}\boldsymbol{f}\b{\tau}d\tau}$, and the elements of the vector $\boldsymbol{f}\b t$ are the corresponding driving terms, $\{f_i\}$, of Eq. \eqref{eq:H_comp1}.

The form of the master equation, \eqref{eq:dot_rho_s_lin}, highlights three phenomena arising from the system-environment interaction: (i) The interaction modifies the primary system's coherent dynamics by effectively coupling different system modes, manifested by the non-diagonal terms of $\boldsymbol{\Omega}$ \footnote{Even for a non-interacting system with a diagonal $M$}, (ii) it leads to the emergence of a drive-dependent coherent term $\boldsymbol{f}_{NM}$, Eq. \eqref{eq:k_t}. For multiple system modes, the non-Markovian correction may lead to a collective effect, where cross-mode coupling elements in $\boldsymbol{W}$ imply that an effective drive on the $i$'th mode, $k_i\b t$, depends on the chosen driving protocols of the other modes $\{f_j\b t\}$. This contribution is weighted in a non-local way by the spectral density function elements, as depicted by Eq. \eqref{eq:f_NM}. (iii) Finally, the form of  $\Gamma_{ij}^{\downarrow,\uparrow}$  highlights that for such a linear system, the incoherent decay rates are independent of the drive.

In the Markovian limit, the Green function obtains the form $\boldsymbol{W}\b t=\exp\b{-i{\cal H}_{eff}t}$, where ${\cal H}_{eff}$ is a constant non-Hermitian matrix (for example, see Subsec. \ref{subsec:dynamical solution}). 
Substituting the exponential form into Eq. \eqref{eq:f_NM} shows that $\boldsymbol{f}_{NM}$ vanishes in this regime. Thus, the coherent correction and collective driving effect are purely non-Markovian. In addition, it constitutes a non-Markovian signature, allowing to  detect deviations from Markovianity by the analysis of the coherent dynamics of linear or weakly excited systems.  

The non-Markovian contributions of $\boldsymbol{f}_{NM}\b t$ scale quadratically with the system-environment coupling. Hence, such a collective driving effect is expected to be significant only for strong system environment-coupling, where the open-system dynamics are highly non-Markovian. In the weak coupling regime, ignoring such open system contributions leads to small coherent errors. Nevertheless, with the ongoing improvement in coherent control and contemporary aspirations for precise control, even small errors may be significant. In order to optimize control fidelity the coherent collective-driving effects can be incorporated in the design of control protocols.

The autonomous representation of the composite system (see beginning of Sec. \ref{sec:framework} and \cite{cohen1998atom}) infers that the source of the coherent correction, $\boldsymbol{f}_{NM}$, is the initial environment state.
Such non-Markovianity is unique, as memory effects commonly arise from the spectral features of the environment and strong system-environment interaction. A similar effect has been witnessed in the non-Markovian evolution of a fermonic open system \cite{windt2024fermionic}. 
The unique source of non-Markovianity highlights that $\boldsymbol{f}_{NM}$ differs qualitatively from the Lamb shift correction to the bare frequencies. The latter originates from the exchange of virtual photons, while the former is associated with the presence of real photons in the field \cite{lamb1947fine,hansch1975doppler}.

\subsection{Beyond the rotating wave approximation}
\label{sec:beyond_RWA}
The standard derivation of $H$, Eq. \eqref{eq:H_comp1}, involves canonical quantization of the analogous classical system. This procedure leads to counter rotating terms proportional to $a_ib_k$ and $a_i^\dagger b_k^\dagger$ \cite{cohen1997photons}. In the optical regime and typical system-environment coupling, these terms contribute rapid fluctuations with an amplitude which is on the order of the coupling, and are therefore usually neglected. Nevertheless, they are essential for the accurate computation of Lamb shifts, dispersive forces and entanglement dynamics \cite{fleming2010rotating}. Moreover, for strong coupling, the counter rotating terms may modify the reduced dynamics substantially \cite{zueco2009qubit,agarwal2012tavis}. 

The presence of counter rotating terms in the composite Hamiltonian breaks the U(1) symmetry of the Hamiltonian in the autonomous representation. Naturally, the reduction in symmetry increases the complexity of the calculation of the reduced system dynamics. 
In the following section we study the structure of the master equation for an initial Gaussian state of the environment and arbitrary quadratic Hamiltonian. For such an initial state, a generalization of Ref. \cite{ferialdi2016exact} leads to the general form. The result highlights that the same qualitative interplay between the drive and dissipation, as deduced in Sec. \ref{subsec:exact_lin_ME}, is maintained beyond the rotating wave approximation. 

The exact non-Markovian master equation, governing the dynamics of a non-driven linear (bosonic) system, coupled to a bosonic environment by a bi-linear interaction term was derived in Ref. \cite{ferialdi2016exact}. In contrast to the previous section, here a general linear coupling is considered: 
$H_I  =\sum_jS_j E_j$, where $S^j$ and $E_j$ are linear combinations of the system and environment creation/annihilation operators, respectively.

The construction presented in Ref. \cite{ferialdi2016exact} starts from the most general completely positive, trace preserving Gaussian map \cite{diosi2014general}. 
By exploiting the initial Gaussian state of the environment and Wick's theorem \cite{wick1950evaluation}, Ferialdi deduced the exact dynamical generator of the reduced system. In the Schr\"odinger picture the generator obtains the form
\begin{equation}
    \f{d}{dt}\rho_S\b t =-i\sb{H_{S0},\bullet}+ {\cal L}^{\text{lin}}\sb{\rho_S\b t}~~,
\end{equation}
with
\begin{multline}
     {\cal L}^{\text{lin}}\sb{\bullet}=   \sum_{jk}\Gamma_{jk}\b t \sb{S_j,\sb{S_k,\bullet}}\\+\Theta_{jk}\b t\sb{S_j\sb{\dot{S}_k,\bullet}}
    - i\Sigma_{jk}\b t\sb{S_j,\{S_k,\bullet\}}\\-i\Upsilon_{jk}\b t\sb{S_j\{\dot{S}_k,\bullet\}}~~.
    \label{eq:diosi_ME}
\end{multline}
Here, $H_{S0}$ is an arbitrary time-independent bosonic quadratic system Hamiltonian, governing the isolated (non-driven) system dynamics. The decay rates: $\Gamma_{jk}$, $\Theta_{jk}$, $\Sigma_{jk}$ and $\Upsilon_{jk}$ include convolutions over memory kernels ${\cal A}_{jk}\b t$, and ${\cal B}_{jk}\b t$, and propagators of the free system ${\cal C}_k^j\b{t-s}$ and $\tilde{\cal C}_k^j\b{t-s}$, see Appendix \ref{apsec:beyond_RWA} for explicit expressions. The propagators of the (non-driven) linear system relate Heisenberg picture operators at different times
\begin{equation}
    S_j\b{s_1} = \sum_{k}{\cal C}_k^j\b{t-s_1}S_k\b t + \tilde{\cal{C}}_k^j\b{t-s_1}\dot{S}_k\b{t}~~.
    \label{eq:320}
\end{equation}

After introducing known results, we next incorporate the influence of a linear drive on the open system dynamics. Consider the following system Hamiltonian 
$$ H_S'\b t = H_{S0} +H_d\b t~~,$$
where  $H_d\b t$ is a linear driving term (same form as in Eq. \eqref{eq:H_comp1}). Transforming to an interaction picture relative to both bare dynamics and the drive (with respect to $H_S'\b t$) leads to a representation where the composite system dynamics are generated by the interaction picture Hamiltonian, 
\begin{equation}
    \tilde{H}_I\b t = \sum_j \tilde{S}_j\b t E_j\b t~~.
\end{equation}
This Hamiltonian is bi-linear in the system-environment bosonic operators and, without loss of generality, can be assumed to satisfy $\text{tr}_E\b{H_I\b t}=0$ \cite{schaller2014open} \footnote{By adding and subtracting a time-dependent scalar to the Hamiltonian one can redefine the system and interaction Hamiltonian terms so to satisfy   $\text{tr}_E\b{H_I\b t}=0$, the details of the derivation are given in Ref. \cite{schaller2014open}}.
As a consequence, one can repeat the construction presented in Ref. \cite{ferialdi2016exact} with only minor modifications.
Foremost, in the presence of a linear drive, the dynamics in the Heisenberg picture acquire an additional scalar term,
\begin{multline}
    \tilde{S}_j\b{s_1} = \sum_{k}{\cal C}_k^j\b{t-s_1}S_k\b t \\+ \tilde{\cal{C}}_k^j\b{t-s_1}\dot{S}_k\b{t}+\bar{{C}}^j\b{t-s_1}~~,
    \label{eq:334}
\end{multline}
where the memory kernels ${\cal C}$ and $\tilde{\cal{C}}$ are identical to the non-driven case, Eq. \eqref{eq:320} and $\bar{{C}}^j$ is a scalar term involving a convolution of the non-driven propagator and the driving term. 
In Appendix \ref{apsec:beyond_RWA}, we show that the modified Heisenberg dynamics, relative to Eq. \eqref{eq:320}, does not alter the resulting decay rates, $\Gamma$, $\Theta$, $\Sigma$ and $\Upsilon$. Its sole contribution is a new coherent term in the master equation, similarly to Eq. \eqref{eq:dot_rho_s_lin}. 

Overall, we obtain general form 
\begin{multline}
    \f{d}{dt}\rho_S\b t  = -i\sb{H'_{S}\b t,\rho_S\b t}\\-i\sum_j\Phi_{j}\b t\sb{S_j,\rho_S\b t}+{\cal L}^{\text{lin}}\sb{\rho_S\b t}~~,
\label{eq:nonRWA_ME}
\end{multline}
where $\Phi\b t$ is given in Appendix \ref{apsec:beyond_RWA}.
This relation provides a more general expression than Eq. \eqref{eq:dot_rho_s_lin}, as it generally includes the dynamical contributions of the counter rotating terms in the system-environment interaction. The downside is that the memory kernels, ${\cal{A}}_{jk}$ and ${\cal B}_{jk}$, 
are given in terms of asymptotic series of complex multidimensional integrals (similarly to the Dyson series). Hence, for practical calculations they require a perturbative analysis. 

Importantly, the structure of the Hamiltonian, Eq. \eqref{eq:nonRWA_ME}, highlights that the inclusion of counter rotating terms does not change the qualitative interplay between coherent and incoherent dynamical contributions. Namely, the coherent drive does not affect the dissipation.
Establishing this point, in the following analysis, we return to the case where the counter rotating terms have been neglected.

\subsection{Exact dynamical solution}
\label{subsec:dynamical solution}
The evolution of the reduced system associated with $H$, Eq. \eqref{eq:H_comp1}, is determined by the non-equilibrium Green function $\boldsymbol{W}$. The Green function is determined by the integro-differential equation  \eqref{eq:W_t} which is independent of the driving terms.  Obtaining analytical solutions to Eq.  \eqref{eq:W_t}, is generally an elaborate task. In the case of a single system  mode and certain memory kernels $\boldsymbol{K}$, a solution can be obtained by employing standard Laplace transform identities \cite{vacchini2010exact}. Beyond these cases, a general methodology applicable to multiple system modes and a broader class of environments is desirable. In the following section, we describe such an approach. By relying on the resolvent formalism \cite{cohen1998atom}, the method provides a systematic way to obtain exact or approximate solutions for $\boldsymbol{W}\b t\equiv \boldsymbol{W}\b{t,0}$. 

The non-driven system dynamics can be concisely expressed in terms of the Heisenberg equations of motion of the operator valued vector $\boldsymbol{v}\b t =\{\boldsymbol{a}\b t,\boldsymbol{b}\b t\}^T$, where $\boldsymbol{a} = \{a_1\dots,a_{N_S}\}^T$ and $\boldsymbol{b}=\{b_1,\dots,b_{N_E}\}^T$, while the time-dependence designates operators in the Heisenberg picture. The composite system dynamics reduce to a Schr\"odinger-like equation 
\begin{equation}
    \dot{\boldsymbol{v}}\b t=-i{\cal H}\boldsymbol{v}\b t~~,
    \label{eq:v_dot}
\end{equation}
where 
\begin{equation}
{\cal H}=\sb{\begin{array}{cc}
\boldsymbol{M} & \boldsymbol{R}\\
\boldsymbol{R}^{\dagger} & \boldsymbol{E}
\end{array}}~~,    
\label{eq:expample_H_cal}
\end{equation}
with $\sb{\boldsymbol{R}}_{il} = \eta_{il}$ and $\boldsymbol{E}=\text{diag}\b{\omega_1,\dots,\omega_{N_E}}$.
Such representation motivates treating the annihilation operators as elements of an $N_S+N_E$-dimensional vector space.

The solution of Eq. \eqref{eq:v_dot} is equivalently stated in terms of the dynamical propagator ${\cal U}\b t=\exp\b{-i {\cal H}t}$. To obtain a solution for $\cal U$, it is beneficial to express the propagator in an integral form
\begin{equation}
    {\cal{U}}\b t=-\f{1}{2\pi i}\int_{-\infty}^{\infty}e^{-iE t}\boldsymbol{G}\b{E+i 0^+}d E~,
    \label{eq:U_int}
\end{equation}
where
\begin{equation}
    \boldsymbol{G}\b{z}=\f{1}{z-{\cal H}}
    \label{eq:resolvent}
\end{equation}
is the resolvent \footnote{The correspondence between  Eq. \eqref{eq:U_int} and its standard exponential form is deduced by an analytical continuation of the integral to the complex plane and application of the residue theorem.}.
The strength of the formalism stems from the form of Eq.  \eqref{eq:U_int}, which enables employing complex contour integration techniques to evaluate the elements of $\cal U$. In addition, Eq.  \eqref{eq:U_int} circumvents the demanding task of exponentiating the large matrix ${\cal H}$, when the inverse, Eq. \eqref{eq:resolvent}, can be accurately evaluated  by alternative techniques.

The form of the propagator naturally decomposes to four block matrices
\begin{equation}
    {\cal U}=\sb{\begin{array}{cc}
\boldsymbol{W} & \boldsymbol{T}\\
\boldsymbol{T}^\dagger & \boldsymbol{P}
\end{array}}~~,
\label{eq:13}
\end{equation}
corresponding to the structure of the resolvent
\begin{equation}
    \boldsymbol{G}\b z=\sb{\begin{array}{cc}
\boldsymbol{G}_{S} & \boldsymbol{G}_{SE}\\
\boldsymbol{G}_{SE}^\dagger & \boldsymbol{G}_{E}
\end{array}}~~,
\label{eq:14}
\end{equation}
where $\boldsymbol{G}_S$ ($\boldsymbol{G}_E$) is a $N_S$ by $N_S$ ($N_E$ by $N_E$) matrix.
The block inversion formula now leads to (see Appendix \ref{apsec:block_inversion}) 
\begin{gather} \boldsymbol{G}_{S}\b z=\boldsymbol{Q}\b z \nonumber\\
    \boldsymbol{G}_{SE}\b z=-\boldsymbol{Q}\b z\boldsymbol{R}\f 1{z-\boldsymbol{E}}  \label{eq:15}\\
\boldsymbol{G}_{E}\b z=\f 1{z-\boldsymbol{E}}+\f 1{z-\boldsymbol{E}}\boldsymbol{R}^{\dagger}\boldsymbol{Q}\b z\boldsymbol{R}\f 1{z-\boldsymbol{E}}~~,\nonumber
\end{gather}
with
\begin{equation}
    \boldsymbol{Q}\b z=\sb{z-\boldsymbol{M}-\boldsymbol{\Sigma}\b z}^{-1}~~,
    \label{eq:16}
\end{equation}
where  the self-energy is given by  
\begin{equation}
\boldsymbol{\Sigma}\b z =    \boldsymbol{R}\f 1{z-\boldsymbol{E}}\boldsymbol{R}^{\dagger}~~.
\label{eq:17}
\end{equation}
Finally, by combining Eq. \eqref{eq:U_int} , \eqref{eq:13}, \eqref{eq:14}, \eqref{eq:15} and \eqref{eq:16}, the non-equilibrium Green function is expressed as
\begin{equation}
    \boldsymbol{W}\b t=
    -\f{1}{2\pi i}\int_{-\infty}^{\infty}\f{e^{-iE t}}{{E^+-\boldsymbol{M}-\boldsymbol{\Sigma}\b{E^+}}}d E~,
    \label{eq:W_int}
\end{equation}
employing the notation  $E^+ = \lim_{\eta\ra 0^+} E+i\eta$.
Similarly, $\boldsymbol{T}$ and $\boldsymbol{P}$ can be expressed in an analogous form, in terms of $\boldsymbol{G}_{SE}$ and $\boldsymbol{G}_E$, respectively. 



For certain environments analytical expressions for $\boldsymbol{\Sigma}$  can be computed analytically \cite{economou2006green,morita1971useful,katsura1971lattice,katsura1971lattice,horiguchi1974lattice,guttmann2010lattice}, numerically \cite{haydock1972electronic,haydock1980recursive,viswanath1994recursion,bulla1998numerical,pulci1998ab,allerdt2019numerically} or perturbatively \cite{cohen1998atom}. 
When an analytical expression for $\boldsymbol{\Sigma}$ is available, the integral of Eq. \eqref{eq:W_int} can be solved by contour integration. The contour integral includes three types of contributions: (a) Real (``stable'') poles, which lead to the non-decaying oscillatory terms of the propagator, correspond to bound system-environment states (b) unstable complex poles, which are manifested in decaying contribution to the propagator and therefore, capture dissipation and (c) branch cuts, which contribute a power-law decay (see Ref. \cite{cohen1998atom} for a detailed analysis).
This enables expressing the Green function as 
\begin{equation}
    \boldsymbol{W}\b t = \sum_{\alpha\in SP}\boldsymbol{R}_\beta e^{-iz_{\alpha}t}+ \sum_{\beta\in UP}\boldsymbol{R}_\beta e^{-i z_{\beta}}+\sum_{\gamma\in BC} C_{\gamma}~~,
\end{equation}
where $SB$ and $UP$ designates the set of stable poles and unstable poles, while $C_\gamma$ are the contributions of the contour segment circumventing the branch cuts.

In the Markovian limit, the weak system-environment coupling and smooth dependence of  density of states on the energy allows approximating  the self-energy by
\begin{multline}
\boldsymbol{\Sigma}\b {E+i0^+} \approx \boldsymbol{\Sigma}\b {E_0+i0^+}=\boldsymbol{\Delta}_0-i\boldsymbol{\Gamma}_0/2~~,    
\end{multline}
where $E_0$ is taken as the average of system energies, while the elements of the Hermitian matrices $\boldsymbol{\Delta}_0$ and $\boldsymbol{\Gamma}_0$ include the collective Lamb shifts and decay rates, Appendix \ref{apsec:markov_dyn}. Such an approximation effectively includes the contribution of only a single pole to the solution of \eqref{eq:U_int}, neglecting the other poles and branch cuts. A straightforward application of the residue theorem now leads to the Markovian solution of the Green function
\begin{equation}
    \boldsymbol{W}\b t\approx\exp\b{-i{\cal H}_{eff} t }~~\\
    \label{eq:W_markov}
\end{equation}
\begin{equation}
    {\cal H}_{eff} = \boldsymbol{M}+\boldsymbol{\Delta}_0-i\boldsymbol{\Gamma}_0/2~~. \nonumber
\end{equation}

Generally, once $\boldsymbol{W}$ is obtained, exactly (non-Markovian regime) or approximately (Markovian or beyond Markovian regime), the extension of the analysis to the case of a driven system is achieved by including an additional
non-homogeneous term in Eq. \eqref{eq:v_dot}. The linear non-homogeneous equation then admits the general dynamical solution
\begin{equation}
    \boldsymbol{v}\b t= {\cal U}\b t \boldsymbol{v}\b 0 -i\int_0^t {\cal U}\b{t-s}\tilde{\boldsymbol{f}}\b s ds~~,
    \label{eq:solution}
\end{equation}
where $\tilde{f}=\{\boldsymbol{f},\boldsymbol{0}\}^T$  is an $N_S+N_E$ size vector, containing the driving terms in its first $N_S$ elements.
Relation Eq. \eqref{eq:solution} provides an exact solution for both the system and environment modes in the presence of a drive. The structure of the second term, and non-diagonal form of $\cal U$, highlights that the system and environment modes are generally affected by the drive.
 
These results conclude the first part of the analysis of linear systems. In the following section, we apply the general relations to study the dynamics of simple primary systems and demonstrate the drive related non-Markovian phenomena.

\subsection{Single harmonic mode coupled to a 1D photonic crystal}
\label{subsec:single harmonic}
We demonstrate the general dynamical solution by considering a minimal model describing a driven nano-cavity coupled to a 1D photonic crystal. The photonic crystal is modeled by a chain of $N_E$ identical bosonic modes, interacting by nearest-neighbour interactions, with periodic boundary conditions and initially in the vacuum state.
The composite system Hamiltonian reads 
\begin{equation}
 H = \omega_S a^\dagger a +g \b{ a b^\dagger_0 + a^\dagger b_0} +H_E^{\b{t.b}}~~,
 \label{eq:H_example}
\end{equation}
where the photonic crystal is represented by
\begin{equation}
    H_E^{\b{t.b}} = \omega_E \sum_{j\in \mathbb{Z}_{N_E}} b_j^\dagger b_j - J \sum_{j\in \mathbb{Z}_{N_E}} \b{b_j^\dagger b_{j+1}+b_{j+1}^\dagger  b_j}~~,
    \label{eq:H_tb_single}
\end{equation}
and the nano-cavity is coupled to single environment site.
The considered model may exhibit strong non-Markovian dynamics, and exotic phenomena as fractional and power-law decay \cite{gonzalez2017markovian}  as well as system-environment bound states \cite{leonforte2021dressed,shi2016bound}, while allowing for an analytical dynamical solution. 

The photonic crystal's bare Hamiltonian can be diagonalized by introducing the  Fourier transformed operators $b_k=\f{1}{\sqrt{L}}\sum_{j} e^{-ik a j}b_j$, with lattice constant $a$ and $k\in\b{2\pi/L}\mathbb{Z}_{N_E}$. The substitution leads to a band structure within the energy range $\sb{\omega_E-2J,\omega_E +2J}$.
In the rotating frame with respect to $\omega_E$, Eq. \eqref{eq:H_tb_single} becomes 
\begin{equation}
    H = \bar{\Delta}  a^\dagger a +g \b{ a b_k^\dagger  + a^\dagger b_k } \\+\sum_k\omega_k b_k^\dagger b_k~~,
    \label{eq:H_example2}
\end{equation}
with $\bar{\Delta} = \omega_S-\omega_E$
The environment spectrum $\omega_k = -2J \cos\b{k a}$ manifests diverging density of states
at the band edges: $D\b E=\f{\theta\b{2J-E}}{\pi\sqrt{4J^{2}-E^{2}}}$ \cite{economou2006green}. 
The rapid change in density of states leads to highly non-Markovian behavior when $|\bar{\Delta}|\approx 2J$ \cite{john1990quantum}, see Fig. \ref{fig:contribution} Panel (b). Interestingly, such increase in the density of states results in an  enhancement in the coupling efficiencies between the system mode and the photonic crystal, which can assist in the realization of single photon transistors, large single photon nonlinearities and creation of single photon sources  \cite{arcari2014near}.

The solutions for $\boldsymbol{W}$ and $\boldsymbol{T}$ follow a procedure similar to that presented in Ref. \cite{gonzalez2017markovian}, studying the spontaneous emission of two-level emitters, coupled to a 1D and 2D cubic lattices. Here, we summarize the results and provide a detailed analysis in Appendix \ref{apsec:two_modes_exp}. Following, we utilize these results to investigate the non-Markovian effects associated with the drive. 
 Equation \eqref{eq:solution} provides the system dynamics
\begin{equation}
a\b t = W\b t a\b 0 + \boldsymbol{T}\b t \boldsymbol{b}\b 0 -i\int_0^t W\b{t-s}f\b{s}~~,
\end{equation}
where in this case the non-equilibrium Green function and the drive term are simply time-dependent scalars 
\begin{equation}
    {W}\b t=
    -\f{1}{2\pi i}\int_{-\infty}^{\infty}\f{e^{-iE t}}{{E^+-{\omega_S}-{\Sigma}_{1D}\b{E^+}}}d E~,
    \label{eq:W_int_single_emit}
\end{equation}
with the self energy \footnote{The expression is directly obtained from taking $n_{12}$ in the expression for $\bar{\Sigma}$ in Appendix \ref{apsec:two_modes_exp}.}
\begin{equation}
    \Sigma_{1D}\b{z} = \f{\text{sign}{\b{\text{Re}\b{z}}}g^2}{\sqrt{z^2-4J^2}}~~.
    \label{eq:Sigma_1D}
\end{equation}
Finally, the integrals of Eq. \eqref{eq:W_int_single_emit}  (similarly for $\boldsymbol{T}$) are solved by contour integral techniques (see \cite{gonzalez2017markovian} and Appendix \ref{apsec:single_emit_1D}). 

 The non-driven dynamics ($f=0$) exhibit similar non-Markovian features as the single excitation case, described in \cite{gonzalez2017markovian}. Fractional decay, for frequencies close to the band edge, arises from an initial overlap with bound states (real poles of $G_S\b z$), and a $t^{-3}$ power-law decay due to the contributions of branch cuts to the integral Eq. \eqref{eq:W_int_single_emit}.

\begin{figure}
    \centering
    \includegraphics[width=8cm]{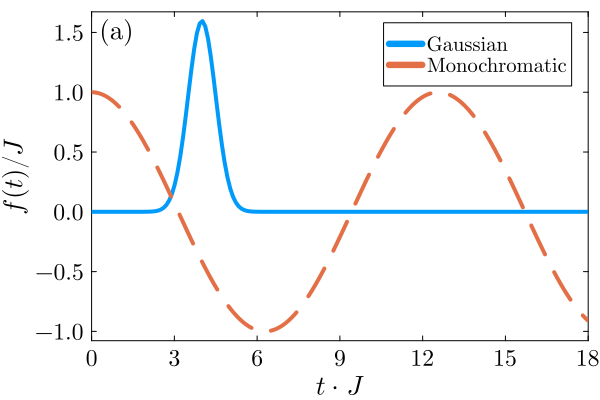}
    \includegraphics[width=8cm]{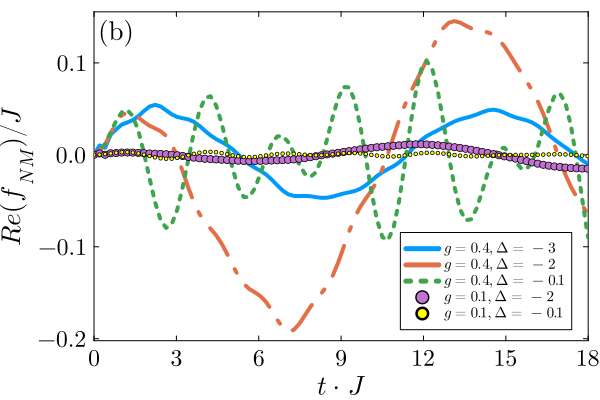}
    \includegraphics[width=8cm]{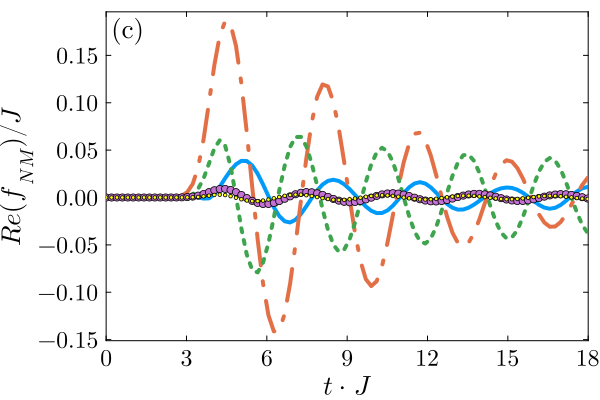}
    \caption{ Non-Markovian influence of the drive. (a) Real part of the control protocols as a function of time. Two types of protocols are analyzed, an off resonant monochromatic drive (continuous blue) $f_{\text{mono}}\b t = \exp\b{-i\b{\omega_S+\delta} t}$, with $\delta =0.5J$  and a Gaussian pulse $f_{\text{gauss}}=(A/\sqrt{2\pi\sigma^2})\b{-(t-t_0)^2/2\sigma_L^2}\exp\b{-i\omega_S t}$. (b,c) The non-Markovian drive dependent coherent correction $f_{NM}$ (Eq. \eqref{eq:f_NM}) as a function of time for a monochromatic drive and Gaussian pulse, respectively. Model parameters: $a=J=1$, $\sigma_L =  0.5 J^{-1}$ and $A=2$.}
    \label{fig:single_mode}
\end{figure}

In the presence of coherent driving, a coherent non-Markovian correction term $f_{NM}$ is added to the master equation \eqref{eq:dot_rho_s_lin}. We analyze the contribution of $f_{NM}$ for two types of control protocols: an off resonant monochromatic drive and a resonant Gaussian pulse, depicted in Fig. \ref{fig:single_mode} Panel (a), for varying coupling strength $g$, $\bar{\Delta}$. We focus on the behavior for negative detunings, as the case of positive detuning show the same qualitative behavior.

At the band edge ($|\bar{\Delta}+2J|\approx g$) for strong coupling ($g=0.4 J$), $f_{NM}$ contributes a shift of up to 10 percent to the coherent drive, depicted by a dashed dot orange line in Fig. \ref{fig:single_mode} Panels (b) and (c). In contrast, for small coupling and frequencies in the band, the magnitude of $f_{NM}$ reduces substantially (yellow small markers). This can be understood by considering the contributions to the contour integration Eq. \eqref{eq:W_int_single_emit}, Fig. \ref{fig:contribution}. 
When the evolution is dominated by a single pole (real or imaginary) the propagator can be approximated as $W\b t\approx c_\alpha e^{-iz_\alpha t}$, where $c_\alpha$ is the residue at pole, which leads to vanishing $f_{NM}$. Therefore, $f_{NM}$ obtains significant values either in the vicinity of the branch cuts or when the contributions of two poles overlap. The branchcut contributions are localized near the band edge, Fig. \ref{fig:contribution} Panel (a), leading to a suppression of the non-Markovian coherent correction well inside the band. Interestingly, in the band gap such a contribution decays slowly, leading to a small but notable contribution even for $|\bar{\Delta} + 2J|/g> 2$ (blue line in Panels (b) and (c) of Fig. \ref{fig:single_mode}). Within the band the contribution of $f_{NM}$ scales quadratically with the coupling strength $g$, as demonstrated by the dashed green line ($g=0.4J$) and yellow markers ($g=0.1J$).

\begin{figure}
    \centering
    \includegraphics[width=8cm]{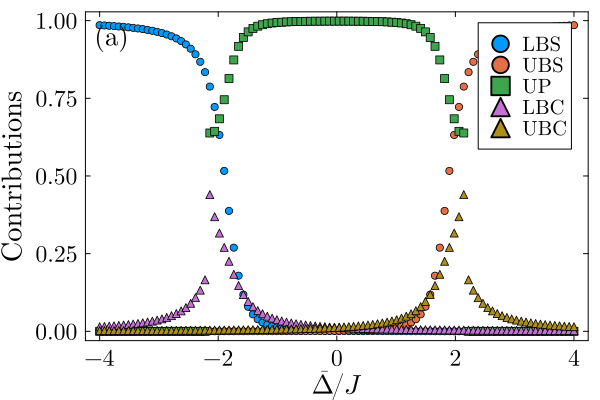}
    \includegraphics[width=8cm]{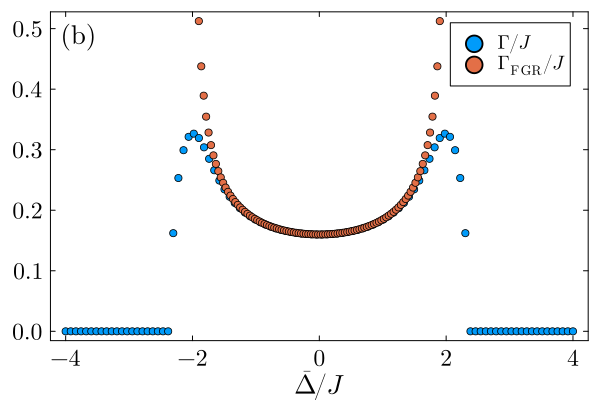}
    \caption{(a) Dynamical contributions to $W\b t$, Eq. \eqref{eq:W_int_single_emit}. There are three types of contributions to the contour integral. (i) Real poles: Lower (RL) and upper (UP). (ii) A single complex pole (CP) and (iii) branch cuts: Lower (LBC) and upper (UBC).(b) Comparison of the single emitter decay exact ($\Gamma$) and  Markovian ($\Gamma_{\text{FGR}} = -\text{Im}\b{\Sigma_{1D}\b{\bar{\Delta}}}/2$) decay rates as a function of the detuning from the photonic crystal's central frequency $\bar{\Delta} = \omega_S-\omega_E$. At the band edges the tight-binding model exhibits a diverging density of states, leading to non-Markovian behavior and a deviation between the Markovian and exact results. Model parameters: $ g=0.4 J$ ($g_{SE}=g$, $J=\Delta \omega_E$).}  
    \label{fig:contribution}
\end{figure}


Comparing the two types of protocols, the non-Markovian contribution of the monochromatic pulse remains of similar magnitude throughout the dynamics, while for the Gaussian pulse the contribution exhibits a peak during the pulse and decays slowly in time. Surprisingly, for the pulse $f_{NM}$ obtains substantial values long after the pulse decays (see dashed-dotted yellow and dashed red line in panel (b)). Such behavior arises from the time non-local nature of the non-Markovian effects.

\subsection{Two driven modes coupled to a 1D photonic crystal}
\label{sec:two_modes}


The following section extends the single-mode analysis to a pair of nano-cavities, coupled to sites $\pm j_{S}$ of a 1D photonic crystal. 
The composite system is represented by the Hamiltonian
\begin{equation}
 H = \omega_S \sum_{n=\pm{j_{{S}}}} a^\dagger_{n} a_n +g\sum_{n=\pm j_{S}} \b{ a_n b^\dagger_n + a_n^\dagger b_n} +H_E^{\b{t.b}}~~.
 \label{eq:H_example_2em}
\end{equation}
Transitioning to the Fourier components decouples the environment's collective modes and leads to an interaction term of the form $g\sum_{n=\pm j_{\text{S}}} \b{ a_n b_k^\dagger e^{i k a n } + a_n^\dagger b_k e^{-i k a n }}$.

\begin{figure}[htb!]
\centering
\includegraphics[width=0.5\textwidth]{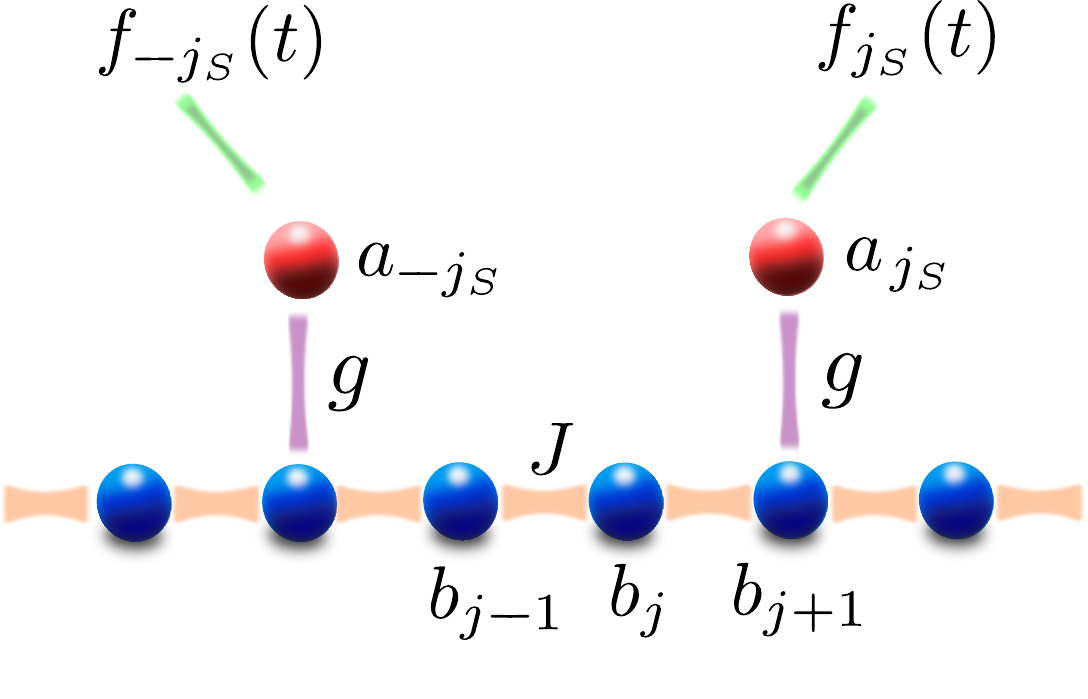}\\
\caption{Schematic representation of the studied linear model. Two driven modes situated at positions $\pm j_S$ with associated drives $f_{\pm j_S}\b t$. The system modes are coupled locally, via a coupling strength $g$, to a 1D photonic crystal, represented by a chain of linear modes with nearest neighbour interactions of strength $J$. The tight-binding environment is characterized by rapidly varying density of states near the band edges, imposing strong non-Markovian system dynamics when $|\bar{\Delta}\pm J|\sim g$. } 
\label{fig:linear_exp_scheme} 
\end{figure}

Reflection symmetry of the configuration enables decoupling the composite system into two collective system modes, $a_{\pm}$, which are coupled to two distinct environments \cite{gonzalez2017markovian}. Thus, effectively reducing the analysis to a single mode problem. 
The system's dynamics take the form  
\begin{multline}
    \boldsymbol{a}\b t= \boldsymbol{W}\b t\boldsymbol{a}\b 0 +\boldsymbol{T}\b t\boldsymbol{b}\b 0~~\\
    -i\int_0^t \boldsymbol{W}\b{t-s}\boldsymbol{f}\b s ds~~,
\end{multline}
where
\begin{equation}
   \boldsymbol{W}\b t= \f 1{{2}}\sb{\begin{array}{cc}
W_{+}+W_{-} & W_{+}-W_{-}\\
W_{+}-W_{-} & W_{+}+W_{-}
\end{array}} 
\label{eq:W_exp}~~,
\end{equation}  
and $W_\pm$ are the non-equilibrium Green function of the collective system modes $a_\pm =\b{a_1\pm a_2}/\sqrt{2}$. These Green functions are given explicitly in Appendix \ref{apsec:two_modes_exp}.

\begin{figure}
    \centering
    \includegraphics[width=8.5cm]{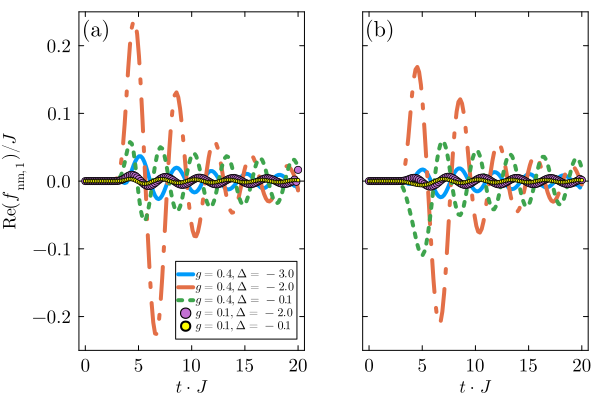}
    \includegraphics[width=8.5cm]{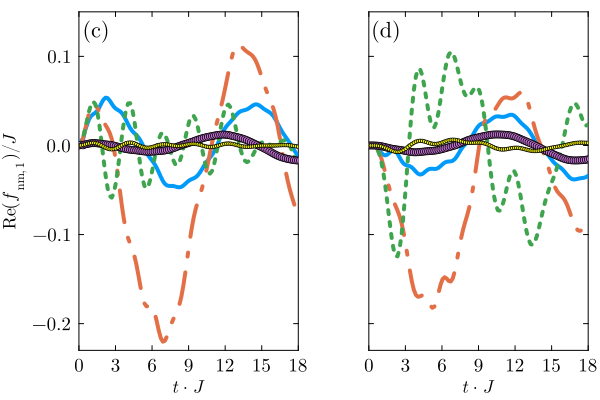}
    \caption{The non-Markovian drive term on the first 
  $f_{NM,1}\b t$ (a,c) and second mode $f_{NM,2}\b t$ (b,d) as a function of time for different system-environment coupling and detunings. The first mode is driven by a Gaussian pulse in (a,b) and monochromatic laser in (c,d).  The model parameters are identical to Fig. \ref{fig:single_mode}.}
    \label{fig:cross_drive_effect}
\end{figure}

\begin{figure}
    \centering
    \includegraphics[width=8.5cm]{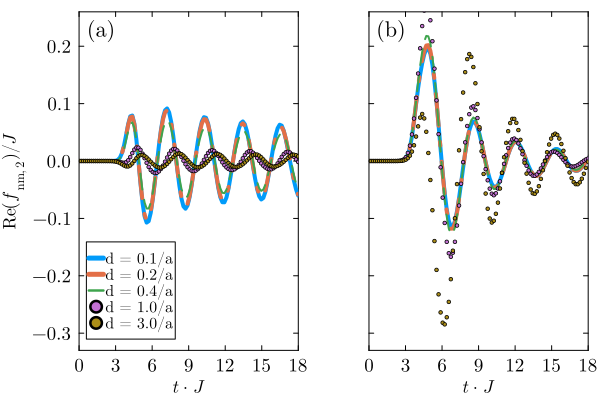}
    \includegraphics[width=8.5cm]{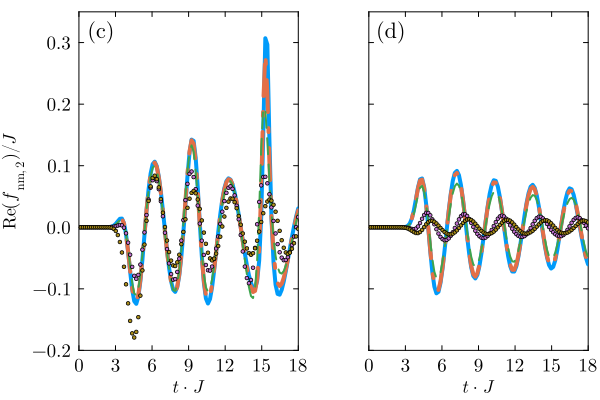}

    \caption{Effect of the distance between the system modes, $d$, on $f_{NM,2}$, for varying detunings and a Gaussian laser pulse on the first system mode. Panels (a,b,c,d) correspond to the detunings $\bar{\Delta} = -3,-2,-1,-0.1$.  $a=J=1$, $\delta = 0.5J$, $\sigma_L =  0.5 J^{-1}$ and $A=2$, $g_{SL} = 0.4 J$.}
    \label{fig:cross_drive_dist}
\end{figure}

The non-diagonal form of $\boldsymbol{W}$ leads to a cross-driving effect, where the drive on one of the system modes influences the other.
The cross-driving effect is demonstrated by considering a  monochromatic drive only on the first system mode ($\boldsymbol{f} \b t=\{A\exp\b{-i \omega_L t},0\}^T $). The drive on the first mode leads to an additional non-Markovian term in the master equation, $f_{NM,2}\b t$,  driving the second mode. Here $f_{NM,i}$ denotes the $i$-th component of the vector $\boldsymbol{f}_{NM}$. The magnitude of non-Markovian cooperative effect increases with the system-environment coupling, Fig. \ref{fig:cross_drive_effect}. The distance between the system modes modifies the collective behaviour and thus, modifies the magnitude of $f_{NM,2}$ substantially. Figure \ref{fig:cross_drive_dist} compares $f_{NM,2}$ for varying distances and emitters detuning $\bar{\Delta}$, for a Gaussian pulse applied to the first emitter.  
Well inside the band and in the band gap the magnitude of the non-Markovian cross effect decreases with the distance (Panels (a) and (d)). This is contrasted by the behavior close to the band edge, where the branchcuts' contribution to the dynamics intensifies. For a plot of the various contributions to the contour integral of $W\b t$ see \cite{gonzalez2017markovian} Fig. (4). For $\bar{\Delta} = -0.1 J,-1 J$ (Panels (b) and (c), respectively) $f_{NM,2}$ exhibits large magnitudes even for large emitter distances. 

\subsection{Linear model validity regime }
The presented analysis and results apply only to linear systems. Nevertheless, since in many cases the non-linearity of physical systems tends to be small \cite{devoret1995quantum}, the analysis can either constitute a good approximation or serve as a basis for a perturbative treatment of the non-linearity.
In addition, when the drive is sufficiently weak such that it generates simultaneously at most a single excitation shared among many particles, the Holstein-Primakoff approximation leads to an effective linear system. For instance, the presented analysis allows treating the dynamics of a weakly driven ensemble of Rubidium atoms in the Rydberg blockade regime \cite{lukin2003colloquium}.   

The analysis is not limited to the optical regime and applies to general driven linear systems coupled to a bosonic environment, where the composite dynamics are governed by a Hamiltonian of the form \eqref{eq:H_comp1}.
For optical frequencies at room temperature, the electromagnetic state is well approximated by a vacuum state. However, when entering the IR regime and below temperature plays a significant role. The temperature effect can be incorporated into the construction following Ref. \cite{yang2013master}. The presence of thermal excitations in the bath modifies the decay rates $\Gamma_{ij}^{\uparrow,\downarrow}$ and the coherent corrections $\Omega_{ij}$, see Appendix \ref{apsec:ME_parameters}. Notably,  the Green function and the non-Markovian driving term $\boldsymbol{f}_{NM}\b t$ is independent of the temperature.

\section{Non-Linear optical systems}
\label{sec:non_linear}

The dynamical behavior of driven linear and non-linear open systems is expected to be qualitatively different. For instance, the perturbative treatment of non-linear systems in the Markovian regime predict drive-dependent dissipative decay rates \cite{kohler1997floquet,albash2012quantum,dann2018time,kuperman2020mechanical,di2024time,mori2023floquet} and a drive-dependent memory kernel in a non-Markovian analysis \cite{meier1999non}. Additionally, the linear case permits a normal mode analysis, while the combination of system non-linearities and the infinite degrees of freedom of the bosonic environment typically prevents exact solution. 
Specifically, the non-linearities imply that the composite Hamiltonian includes quadratic or higher-order terms. As a result, the operator algebra of the composite system leads to an infinite hierarchy of coupled linear differential equations (the Heisenberg equations) \cite{bruus2004many}.
Despite of this fact, for a large class of non-linear systems, the driven open system dynamics can be connected to the dynamics of associated linear systems.


We focus on the driven open system dynamics of an ensemble of two-level quantum emitters. The composite Hamiltonian is of the form
\begin{multline} H\b t=\sum_{ij}\sigma_{i}^{\dagger}M_{ij}\sigma_{j}+\sum_i\b{f_i\b t\sigma_i^\dagger +f_i^{*}\b t\sigma_i}\\+\sum_{ik}\b{g_{ik}\sigma_{i}^{\dagger}b_{k}+g_{ik}^{*}\sigma_{i}b_{k}^{\dagger}}+\sum_{k}\omega_{k}b_{k}^{\dagger}b_{k}~~,
\label{eq:692}
\end{multline}
where $\sigma_i = \ket{g}_i\bra{e}_i$ is the ladder operator of the $i$'th emitter and the environment is assumed to be initially in the vacuum state.
Within the Markovian regime, dynamics of a variety of the optical, solid-state and molecular systems are well represented by Eq. \eqref{eq:692} \cite{barreiro2011open,wang2024simulating,doherty2013nitrogen,goldman2015phonon,pachon2011physical,ishizaki2012quantum,blais2021circuit}. Beyond the Markovian regime, such an Hamiltonian may be realized by neutral and solid-state atoms interacting
with photonic modes confined to engineered dielectric materials \cite{lodahl2015interfacing,gonzalez2017markovian}, cold atoms in a state-dependent optical lattice \cite{de2008matter,navarrete2011simulating} and cavity QED setups \cite{kimble1998strong,walther2006cavity}.

The structure of Eq. \eqref{eq:692} is the same as the linear system Hamiltonian, Eq. \eqref{eq:H_comp1}. This connection enables harnessing the  analytical linear solution to deduce a reliable dynamical solution for the studied non-linear system. The solution  manifests a dynamical equation, termed the ``{\textit{Linear Master Equation}}'' (LME). 
In the following sections we compare the prediction of the LME's to an exact pseudo-mode solution \cite{tamascelli2018nonperturbative,garraway1997decay} and analyze the interplay between the drive and dissipation.

We begin by describing this technique in Sec. \ref{subsec:pseudo}. Following, in Sec. \ref{sec:lin_ME} we map the non-Markovian bosonic equation \eqref{eq:dot_rho_s_lin} to  the LME, a master equation for the emitters, and benchmark its performance in the following sections.  The interplay between the coherent and incoherent dynamical contributions is studied in sections \ref{subsec:strong_driving_single_emiter}, \ref{subsubsection:bandwidth}, \ref{sec:markov_regime}, while the drive dependent cooperative effects are investigated in Sec. \ref{subsec:two_emitters}.
Concluding the analysis of non-linear systems,  Sec. \ref{sec:markov_regime} focuses on the standard optical regime, involving moderate driving strengths and a Markovian environment. In this regime, the accuracy of the linear  and optical Bloch master equations is compared to the Floquet \cite{ho1986floquet,kohler1997floquet,breuer1997dissipative,chu2004beyond,szczygielski2013markovian,szczygielski2014application,elouard2020thermodynamics}, Adiabatic \cite{albash2012quantum} and Time-Dependent \cite{di2024time} master equations.




\subsection{Pseudo-mode solution}
\label{subsec:pseudo}
Simulation of the bosonic environment with dissipative pseudo-modes enables the evaluation of the exact driven open system dynamics. The simulation allows replacing an infinite number of modes with a finite number of so-called pseudo-modes, which are bosonic modes interacting with the system while simultaneously undergoing Markovian dissipation. Such an exact mapping is possible under the following conditions \cite{tamascelli2018nonperturbative,garraway1997nonperturbative}: (a) The environment is bosonic, (b) it is initially in a Gaussian state, (c) the bare environment Hamiltonian is quadratic in the bosonic creation and annihilation operators, and (d) the spectral density of the environment can be well approximated in terms of a linear combination of Lorentzian functions. 

Underlying the mapping are two key insights. First, the influence of the environment on the open system is completely determined by the system interaction operators in the interaction picture and the bare environment's correlation functions \footnote{This is a direct consequence of the Dyson series.}. In addition, for a Gaussian initial state, Wick's theorem implies that high-order correlation functions can be reduced to products of the first and second-order correlation functions. As a result, a bosonic field can be simulated by a reservoir including a finite number of modes, if the environment's and reservoir's first and second correlation functions, as well as the interaction operators in the interaction picture, coincide.
Tamascelli et al. \cite{tamascelli2018nonperturbative} provided a rigorous proof for the open system simulation of a bosonic Gaussian environment, focusing on a non-driven system. Nevertheless, as pointed out in Ref. \cite{pleasance2020generalized}, the proof remains valid for any system Hamiltonian, including arbitrary time-dependent terms.

We consider identical coupling of the emitters to the environment, $g_{ik}=g_k$ (Eq. \eqref{eq:692}), a Lorentzian spectral density function
\begin{equation}
J\b \omega = \f{1}{\pi}\f{g_{SE}^2 \kappa }{\b{\omega-\eta}^2 +\kappa^2}~~,
   \label{eq:lorentz_spect}
\end{equation}
and an environment initially in the vacuum state. The influence of the environment on the studied system can be precisely simulated by a reservoir, $R$, including a single dissipative pseudo-mode. The composite system-reservoir dynamics are given by  
\begin{equation}
    \f{d}{dt}\rho_{SR}\b t = -i\sb{H_{SR}\b t,\rho_S\b t} +2\kappa {\cal D}_{c,c^\dagger}\sb{\rho_{SR}\b t}~~,
    \label{eq:9171}
\end{equation}
where $c$ is the annihilation operator of the pseudo-mode and the composite Hamiltonian reads
\begin{equation}
H_{SR}\b t = H_S\b t +\lam \b{Sc^\dagger +S^\dagger r} +\eta c^\dagger c~~,
\end{equation}
with $H_S \b t = \sum_{ij}\sigma_i^\dagger M_{ij}\sigma_j +\sum_{i}\b{f_i\b t \sigma_i^\dagger+f^{*}\b t\sigma_i}$ and $S =\sum_i \sigma_i$.
The equivalence between the reduced dynamics, $\rho_S\b t$, as obtained from the unitary description Eq. \eqref{eq:692}, and the dissipative system $\text{tr}_R\b{\rho_{SR}\b t}$ Eq. \eqref{eq:9171}, is guaranteed by the equivalence of the interaction operators $S\b t =U_S^\dagger \b t S U_S\b t$, where $U_S\b t = {\cal T}\exp\b{-i\int_0^t d\tau H_S\b{\tau}}$, and the second order correlation functions 
\begin{multline}
    C_E\b{t,0} = \text{tr}_E\sb{\b{\sum_{k}g_{k}b_ke^{-i\omega_k t}}\b{\sum_{k}g_{k}^{*}b_k^\dagger}\rho_E\b 0}\\ = \int_0^\infty J\b{\omega}e^{-i\omega \tau}d\tau\approx {g_{SE}}^2 e^{-\kappa \tau}\\=\text{tr}_R\b{c\b t c^\dagger\b 0 \rho_R\b 0}=C_{R}\b{t,0}~~.
\end{multline}
Here, both $\rho_E\b 0$ and $\rho_R\b 0$ are the respective vacuum states and the dynamics of $c\b t=e^{-i\b{\eta-i\kappa/2 }t}c\b 0$ is evaluated with the adjoint master equation of Eq. \eqref{eq:9171}. The approximation in the third line involves taking the bottom range of the integral to $-\infty$. This is well justified when the Lorentzian decays sufficiently fast $\kappa\ll \eta$.
Physically, the approximation amounts to discarding the expected power-law decay at asymptotically large times \cite{knight1978non}.

To evaluate the system dynamics, the infinite states energy ladder of the pseudo-mode is truncated and the composite system-reservoir state is propagated with a standard numerical propagator.
The number of Fock states required in the calculation can be estimated as $n_{\text{trunc}}> \b{g_{SE}g_{SL}}/\b{\kappa\sqrt{g_{SE}^2+g_{SL}^2}}$.
The validity of the calculation was confirmed by comparing the solution to exact analytical results for non-driven systems \cite{jaynes1963comparison,vacchini2010exact,xia2024markovian} and by verifying that the amplitude of the highest pseudo-mode eigenstates remains negligible throughout the evolution.


\subsection{Linear master equation}
\label{sec:lin_ME}

The bosonic master equation \eqref{eq:dot_rho_s_lin} provides a basis for an approximate equation of motion for the driven emitters' open system dynamics. We assume the back-reaction of the environment is only negligibly affected by the non-linearity of the system, and map the bosonic operators to emitters' ladder operators $b_i\ra \sigma_i$ 
\begin{multline}
    \dot{\rho}_S\b t = - i\sb{\tilde{H}_S\b t,\rho_S\b t}\\ +\sum_{i,j=1}^{N_S}\Gamma_{ij}^{\downarrow}\b t{\cal D}_{\sigma_j,\sigma_i^\dagger}\sb{\rho_S\b t} + \Gamma_{ij}^{\uparrow} \b t {\cal D}_{\sigma_i^\dagger,\sigma_j}\sb{\rho_S\b t}~~,
    \label{eq:lin_ME}
\end{multline}
where ${\cal D}_{x,y}\sb{\bullet}\equiv x\bullet y-\{yx,\bullet\}/2$, $N_S$ is the number of emitters 
\begin{equation}
    \tilde{H}_S\b t = \sum_{i,j=1}^{N_S} \sigma^{\dagger}_i{\Omega}_{ij}\b t \sigma_j+\sum_{i=1}^{N_S}\b{k_{i}\b t \sigma_{i}^{\dagger}+k_i^*\b t\sigma_i}~~.
\end{equation} 
Here, the  relaxation rates and coherent term  are identical to those in the bosonic equation, given explicitly in Appendix \ref{apsec:ME_parameters}. 
Equation \eqref{eq:lin_ME} defines the general form of the linear master equation (LME).

The conducted mapping can be understood in terms of the Holstein-Primakoff transformation \cite{holstein1940field}, which enables expressing spin operators in terms of bosonic operators. The substitution is the zeroth order of such a transformation and therefore is expected to be valid only in the low excitation regime \cite{porras2008collective}. 
Nevertheless, we find that Eq. \eqref{eq:lin_ME} provides an accurate description of the reduced system dynamics beyond the low excitation regime. Finally, in the Markovian limit, the normalized frequencies and decay rates converge to the Markovian values described in Sec. \eqref{sec:linear_sys}, and the linear equation coincides with the well-known optical Bloch master equation \cite{scully1999quantum}.

\subsection{Single emitter Markovian regime}
\label{subsec:strong_driving_single_emiter}

For a single driven emitter coupled to a bosonic environment, Eq. \eqref{eq:692} simplifies to
\begin{multline}
    H\b t =  \f{\omega_S}{2}\sigma_z+f\b t\sigma_+ +f^*\b t\sigma_-\\ +\sum_k\b{g_k \sigma_+ b_k +g_k^*\sigma_- b_k^\dagger}+\sum_k \omega_k b_k^\dagger b_k~~.
    \label{eq:938}
\end{multline}
When the environment is initially in the vacuum state with a Lorentzian spectral density, Eq. \eqref{eq:lorentz_spect}, the emitter's reduced dynamics can be simulated via an artificial environment consisting of a single `pseudo'-mode ($r$) undergoing dissipative dynamics \cite{garraway1997decay,tamascelli2018nonperturbative}. The pseudo-mode's Hilbert space is truncated, and the composite emitter-pseudo-mode density matrix dynamics are evaluated numerically.


 \begin{figure}
    \centering
    \includegraphics[width=8cm]{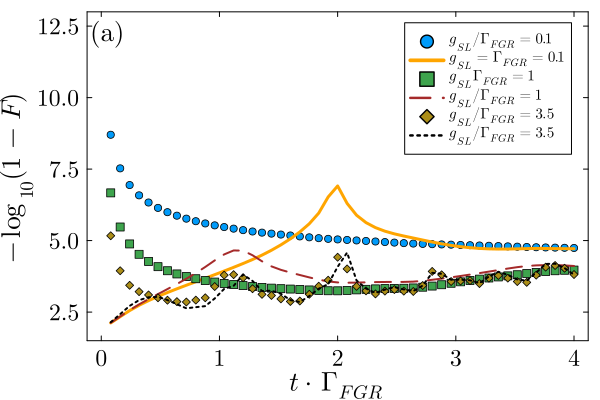}\\
    \includegraphics[width=8cm]
    {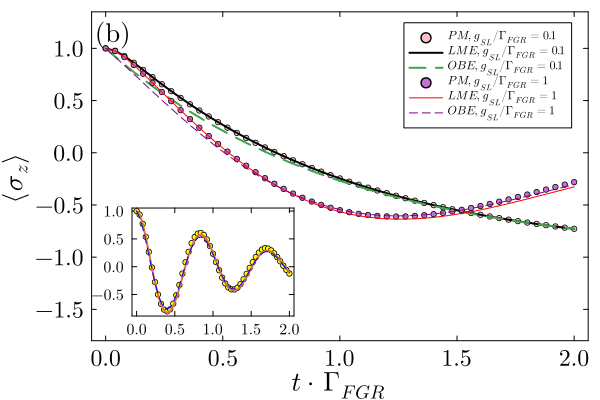}
    \caption{Monochromatic drive in the Markovian regime: Comparison of the LME (markers) and OBE (thin lines) for increasing driving strength. (a) Negative logarithm of the infidelity with respect to the pseudo-mode solution as a function of time. (b) Dynamics of  $\mean{\sigma_z\b t}$. Inset presents the PM (yellow circles), LME (continuous orange line) and OBE (dashed blue line) for $g_{SL} = 3.6 \Gamma_{FGR}$. The driving strength is large relative to the chosen detuning $\Delta = \omega_S-\omega_L = 10^{-4}\omega_S$; for weaker driving the agreement between the LME and the PM solutions improves. Model parameters: $\ket{\psi_S\b 0}=\ket{e}$, $\Gamma_{FGR} = 2\pi g_{SE}^2 D\b{\omega_S} \approx 0.011\omega_S \approx 0.055 \Delta \omega_E
    $, $\Delta \omega_E = \kappa = 0.2 \omega_S$. The point at time $t=0$, where both solutions are exact, is excluded from all the infidelity comparisons. }
\label{fig:sigma_z_mono_non_Markov}
\end{figure}

 \begin{figure}
    \centering
    \includegraphics[width=8cm]{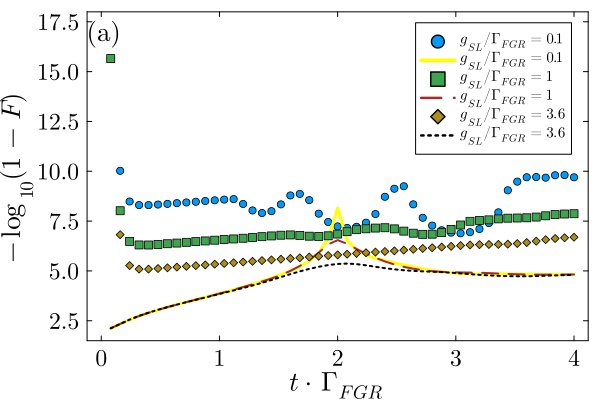}\\
    \includegraphics[width=8cm]
{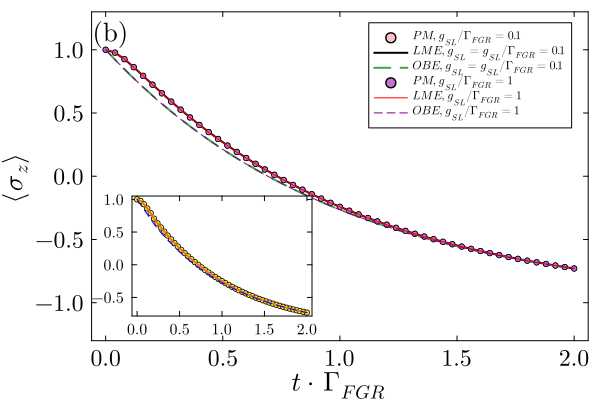}
    \caption{
    Gaussian pulse in the Markovian regime: Comparison of the performance of the LME (markers) and OBE (thin lines) for increasing driving strength. (a) Negative logarithm of the infidelity with respect to the pseudo-mode solution as a function of time. (b) $\sigma_z$ expectation value as a function of time. Inset: identical coloring as in Fig. \ref{fig:sigma_z_mono_non_Markov}, $g_{SL} = 0.1 \omega_S=3.6 \Gamma_{FGR}$. Model parameters: $t_0 = 15/\omega_S =0.17 \Gamma_{FGR}$, $\sigma_L \approx 2 /\omega_S^{-1}=0.02\Gamma_{FGR}$, while the rest of the parameters are identical to Fig. \ref{fig:sigma_z_mono_non_Markov}. }
\label{fig:gauss_Markov}
\end{figure} 


We compare the pseudo-mode solution to the linear master equation (LME) (Eq. \eqref{eq:lin_ME})
\begin{equation}
    \dot{\rho}_S\b t =-i\sb{ H_S^{\b{LME}}\b t,\rho_S\b t}+\Gamma^{\downarrow}\b t D_{\sigma_-,\sigma_+}\sb{\rho_S\b t}~~,
\end{equation}
with $ H_S^{\b{LME}} \b t=\Omega\b t \sigma_+ \sigma_- +k\b t \sigma_+ +k^*\b t\sigma_-$. For this simple model the normalized frequency and decay rates reduce to $\Gamma^{\downarrow}\b t = -\b{A+A^*}$, $\Omega\b t =i\b{A-A^*}/2$ with $A\b t=\dot{W}/W$. The non-equilibrium Green function can be obtained from Eq. \eqref{eq:W_t} by a Laplace transform 
\begin{equation}
    W\b t = e^{-\b{i\omega_S-\kappa/2}t}\b{\cosh\b{d t}+\f{\kappa}{2d}\sinh\b{dt}}~~,
\end{equation}
where $d=\sqrt{\b{\kappa/2}^2-|g_{SE}|^2}$, for a  Lorentzian spectral density, centered at $\eta=\omega_S$.
In the Markovian limit, $\Gamma^\downarrow\b t\ra \Gamma_{FGR}= 2\pi J\b{\omega_S}$, $\Omega\ra\omega_S$, $k\b t \ra f\b t$, and the LME converges to the Optical Bloch Master Equation (OBE)
\begin{equation}
    {\cal L}^{\b{OBE}}\sb{\bullet} = -i\sb{{H}_{S}^{\b{OBE}}\b t,\bullet} +\Gamma_{FGR}{\cal D}_{\sigma_-,\sigma_+}\sb{\bullet}~~,
\end{equation}
with ${H}_S^{\b{OBE}}\b t = \b{\omega_S/2}\sigma_z +g_{SL}\b{f\b t\sigma_+ + f^*\b t \sigma_-}$.


We start by analyzing the Markovian regime. Figure \ref{fig:sigma_z_mono_non_Markov} benchmarks the LME and OBE master equations with respect to the PM solution, for increased driving strength. 
At short times ($t<1/\Gamma_{FGR}$) the LME closely matches the exact solution achieving fidelities as low as $F\approx 10^{-6}$ for $g_{SL}\sim 0.1 \Gamma_{FGR} $. Increasing the driving strength leads to a decrease in accuracy (characterized by $-\log_{10}\b{1-F}$) due to the non-linear effects, reaching infidelities of $1-F\sim 10^{-3}$. At intermediate times the accuracy of the OBE surpasses the LME while converging to similar accuracy at long times, Panel (a). Panel (b) showcases the typical dynamics of an emitter expectation value. Remarkably, the LME correctly captures the non-Markovian initialization step. Smoothly connecting the short-time quasi-unitary dynamics to the long-time exponential decay predicted by the OBE.
For short times, the discrepancy between OBE and the exact result arises from finite information flow speed. Alternatively, the decay is quadratic in time $\mean{\sigma_z\b t}\sim\exp\b{-\b{g_{SE}t}^2}$ as $t\ra 0$ \footnote{It can be derived by applying time-dependent perturbation theory to the Heisenberg equations of motion.}.  The large deviation between the LME and OBE, e.g. Fig. \ref{fig:sigma_z_mono_non_Markov} Panel (a) demonstrates that non-Markovianity may lead significant differences in accuracy, at times much larger than expected $\gg 1/\Delta \omega_E$ ($\sim 0.05/\Gamma_{FGR}$ in the in the model parameters of Fig. \ref{fig:sigma_z_mono_non_Markov}).

The Gaussian pulse exhibits an improved precision, as shown in Fig. \ref{fig:gauss_Markov}. Before the pulse, the LME agrees with the PM to within numerical precision, while the OBE exhibits infidelities of the order $10^{-2}$. The pulse creates small deviations with respect to the exact result, obtaining an improvement of between one and two order of magnitude relative to the monochromatic laser. Such an improvement results from the restricted time-range of the drive. This reduces the population transfer between the ground and excited states and consequently the non-linear open system effects. The fidelity of the LME increases at long times, maintaining significant improvement in accuracy relative to the OBE prediction. This typical behavior differs qualitatively from the monochromatic laser case. Note that even for the strongest studied driving strength, the departure from the exact solution is unnoticeable in the emitter observable dynamics, e.g., inset of Panel (b).

\subsection{Narrow pulse bandwidth non-Markovian effect}
\label{subsubsection:bandwidth}

For sufficiently short pulses, the laser bandwidth becomes comparable to the environment's bandwidth. The competition between these timescales is expected to lead to novel dynamical effects \cite{kruchinin2019non}. 
This regime was explored by comparing the evolution of $\mean{\sigma_z\b t}$ as predicted by the LME, OBE, and PM solutions for varying pulse standard deviation, $\sigma_L$, of a normalized Gaussian pulse

\begin{figure}
    \centering
    \includegraphics[width=8cm]{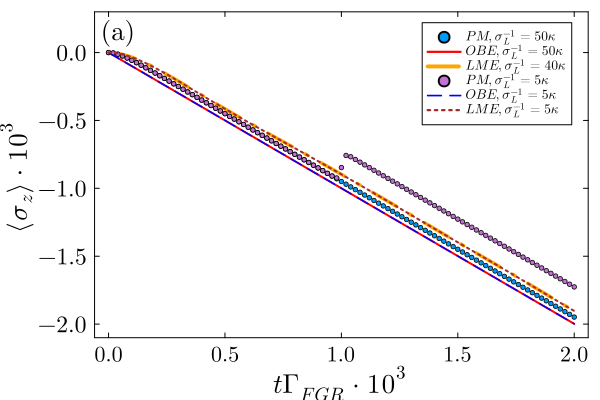}
    \includegraphics[width=8cm]{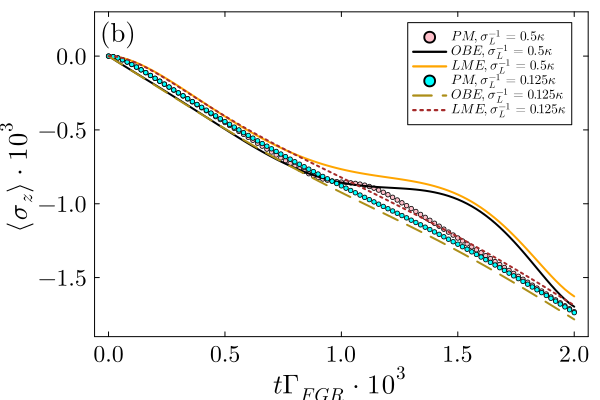}
    \caption{Pseudo-mode, Linear and Optical Bloch master equation solutions for varying pulse width, $\sigma_L$. Panels (a) and (b) show the dynamics for fast (wide-bandwidth) and slow (narrow-bandwidth) pulses relative to the environment bandwidth ($\Delta\omega_E=0.2 \omega_S$).   
   Model parameters: $\Delta = 10^{-5}\omega_S$, $\Gamma_{FGR} \approx 5\cdot 10^{-5} \Delta \omega_E$, $\Delta \omega_E = \kappa = 0.2\omega_S$, $\eta = \omega_S$, $t_0 = 100/\omega_S$ and an initial composite state $\ket{\psi\b 0} = \sb{\b{\ket{g}+\ket{e}}/{\sqrt 2}} \otimes \ket{\text{vac}}$.}
    \label{fig:sigma_L_comp}
\end{figure} 

The qualitative behavior depends on the ratio between the pulse and environment bandwidths ($\Delta \omega_E =\kappa= 0.2 \omega_S$).  
For an ultra-short pulse ($\sigma_L^{-1}=50 \kappa$, Fig. \ref{fig:sigma_L_comp}, Panel (a)) the pulse is much faster than the fastest timescale of the composite system in the rotating frame ($\kappa$). As a result, the sudden approximation holds, and the system and environment cannot respond to the rapid change in the Hamiltonian \cite{sakurai2020modern} \footnote{A similar qualitative effect was found in Ref. \cite{meier1999non}.}. When the  pulse's bandwidth  is narrower, yet within the environment's range ($\sigma_L^{-1} = 5 \kappa$, green circles), the pulse causes a rapid shift in the expectation value, which is not captured by both the LME and OBE solutions. 
Decreasing the pulse's bandwidth below the environment's smooths out the rapid shift (Panel (b), pink circles). This behavior is only partially captured by the LME and OBE (orange and black continuous lines, respectively), which overestimate the feature. For slower pulses relative to $\Delta \omega_E$ (narrow bandwidth), the shift in $\mean{\sigma_z}$ is smoothed out further (cyan circles), which is accurately predicted by both the LME and OBE solutions.

To conclude, for sufficiently rapid driving (in the rotating frame), where the pulse and environment bandwidths become comparable, non-Markovian effects may give rise to deviations between the LME (OBE) and the exact solution.  In the ongoing efforts to achieve faster control protocols and combat inevitable dissipation processes, it is expected that even in experiments done in free space precise control will eventually require accounting for such non-Markovian effects. 
Alternatively, in highly non-Markovian environments, such as high-Q cavity QED, even narrow-band pulses can match the cavity decay rate $1/\tau_{\text{decay}}\sim\Delta\omega_E$.
In this regime, narrow-bandwidth non-Markovian effects could be demonstrated in tabletop cavity QED experiments

\subsection{Non Markovian regime}
\label{subsec:non_markov_regime}

\begin{figure}
    \centering
    \includegraphics[width=8cm]{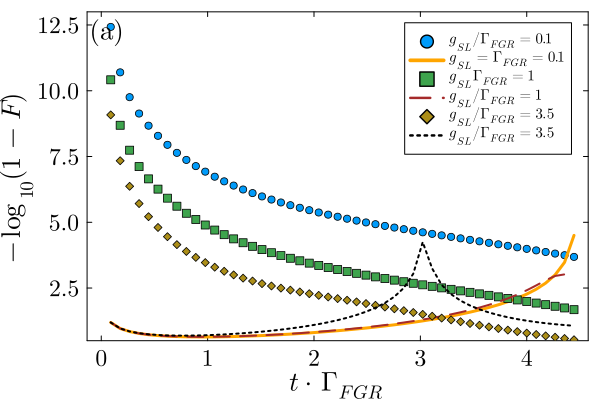}
    \includegraphics[width=8cm]{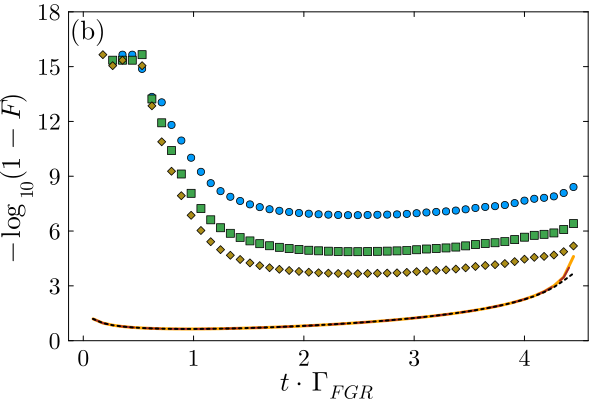}
    \includegraphics[width=8cm]{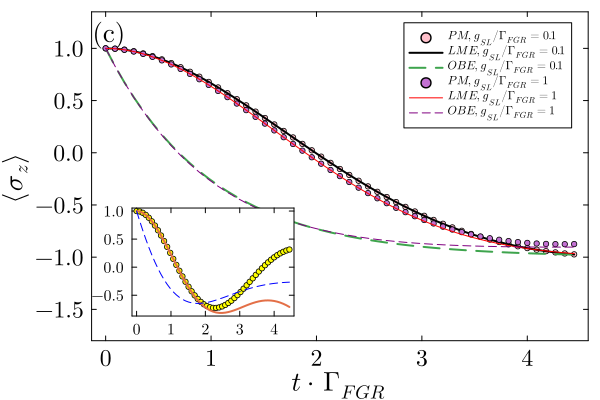}
    \caption{ Non-markovian dynamics: Infidelity (a,b) and $\mean{\sigma_z\b t}$ as a function of time for a monochromatic laser (a,c) and Gaussian pulse (b). Inset: comparison of the LME, OBE and PS solutions for $g_{SL}/\Gamma_{FGR}=3.5$.   Model parameters: $\Gamma_{FGR}/\Delta \omega_E = 0.4$, $\Delta \omega_E = 0.03\omega_S$ and $\ket{\psi_S\b 0}=\ket{e}$. The Gaussian pulse is centered at $t_0\approx 1.1/\Gamma_{FGR}$ with a root mean square of $\sigma_L \approx 0.15 \Gamma_{FGR}$.  }
\label{fig:infidelity_non_Markov}
\end{figure}

When the linewidth ($\Gamma_{FGR}$) becomes comparable to the environment bandwidth, the non-Markovian become significant. The LME then provides an accurate solution of the dynamics for the short and intermediate time regimes and moderate driving strength, Fig. \ref{fig:infidelity_non_Markov} Panel (a). For a monochromatic drive, the accuracy of the LME drops sharply due to the non-linear corrections, while for a Gaussian pulse, the solution remains accurate even at long times, Panel (b). Panel (c) compares the non-Markovian (LME) and Markovian (OBE) evolution of $\mean{\sigma_z\b t}$. As depicted in the figure, the non-Markovianity modifies the solution substantially.

\subsection{Drive dependent dissipation: comparison to the Adiabatic, Floquet and Time-dependent master equations}
\label{sec:markov_regime}



 
A central question in the study of driven open systems is: Does the coherent drive affect the dissipation rates? 
To address this issue, we benchmark the performance of the Adiabatic (AME) \cite{albash2012quantum}, Floquet (FME) and Time-Dependent (TDME) \cite{di2024time} master equations, which incorporate the influence of the drive on the dissipation rates. The performance of these master equations are compared to the LME and the OBE which are characterized decay rates and Lindblad jump operators which are independent of the drive.



In the interaction picture relative to the central laser frequency, $\omega_L$, the AME, FME and TDME are of the form
\begin{equation}
    {\cal L}_{\text{A}}  = -i\sb{\tilde{H}_{S}\b t,\bullet} +{\cal D}^{\b{\text{A}}}\sb{\bullet}~~,
    \label{eq:L^A}
\end{equation}
 with distinct dissipators $\text{A}=\text{AME, FME, TDME,}$ given explicitly in Appendix \ref{apsec:Markov_ME}.
Notably, in order to accurately evaluate the Markovian decay rates of ${\cal D}^{\b{A}}$ one must account for the shift in the environment's spectrum and density of states, due to the transition to the rotating frame. In the rotating frame, the spectrum range is given by $\Delta_k\in [-\omega_L,\infty)$, and the
spectral density function  becomes $\tilde{J}\b{\omega} = \b{g_{SE}^2\kappa/\pi}/\sb{\b{\omega-\b{\eta-\omega_L}}^2+\kappa^2}$. 


For a monochromatic drive, since we are working in a rotated frame $\tilde{H}_S$ is independent of time and the adiabatic theorem is exact.  
In addition, in such a frame the TDME coincides with the AME. Meaning that both master equations have decay rates and jump operators which depend on the drive, describing the dissipation in terms of transitions between the system's dressed states.

Here we summarize the main findings, while the corresponding Figure (\ref{fig:infidelity_Markov}) is presented in Appendix \ref{apsec:driven_open_non_lin}.
In the weak driving regime, $\Delta \gg g_{SL}$, the FME and AME converge to the OBE, where $\Delta =\omega_S-\omega_L$ is the detuning with respect to the central laser frequency. In contrast, for comparable $\Delta \sim g_{SL}$ or small $\Delta \ll g_{SL}$ detuning, the FME and AME deviate from the exact result. For a monochromatic  drive (Gaussian pulse) the FME, AME and TDME achieve an accuracy in the range $-\log\b{1-F}=0.1-2$ (0.1-4) while the LME and OBE showcase a substantially improved infidelities $F=10^{-3}-10^{-6}$ ($10^{-3}-10^{-10}$),  for varying driving strengths.

Overall, the FME, AME, and TDME do not accurately capture dissipation within the studied parameter regime, suggesting that interpreting dissipation in terms of transitions between Floquet or dressed states may be flawed.


\subsection{Two emitters Markovian and non-Markovian regimes}
 \label{subsec:two_emitters}

The coupling of multiple emitters to common environmental modes enables analyzing the influence of the non-linearities and memory effects on the collective behavior. 
The LME exhibits a good agreement with the exact solution for both short and long times in the Markovian regime, and up to intermediate times in the non-Markovian regime.  The relative accuracy enables understanding and interpreting the driven open system dynamics by means of the different terms of the master equation.

The form of the  LME is given in Eq. \eqref{eq:lin_ME}, with its coefficients determined by the non-equilibrium Green ``function'' $\boldsymbol{W}\b t$. For the studied model, $\boldsymbol{W}\b t$ can be calculated employing the Heisenberg equations of motion of the associated linear system (replacing the ladder operators by bosonic creation annihilation operators). For two emitters with an identical detuning $\Delta$, an explicit analytical form for the Green function is readily obtained. It corresponds to the upper left two-by-two matrix $\exp\b{-i {\cal M}t}$, where $\cal{M}$ is defined by the Heisenberg equation of motion $\dot{\boldsymbol{v}}=-i{\cal M}{\boldsymbol{v}}$, with ${\boldsymbol{v}} = \{a_1,a_2,r\}^T$, see Appendix \ref{apsec:two_emit} for further details.

In the Markovian limit, the LME converges to the OBE, which reads (in the rotated frame)
\footnote{For the simple two emitter case the OBE can be derived by writing the Heisenberg equations of motion for $\sigma_{-}$ and a general field mode $b_k$ and applying the Wigner-Weisskopf approximation which leads to the Markovian observables 
 dynamical equations. Finally,  comparing coefficients of the two equations sets the Lamb shifts and decay rates of the adjoint master equation of Eq.  \eqref{eq:collec_OBE}}
\begin{multline}
    \f{d}{dt}\rho_S\b t=-i\sb{H_S\b t +\sum_{n,m=1,2}J_{nm}^{\b{OBE}}\sigma_m^\dagger \sigma_n}\\+\sum_{nm}\Gamma_{nm}^{\b{OBE}}{\cal D}_{\sigma_m^\dagger,\sigma_n}\sb{\rho_S\b t}~~.
    \label{eq:collec_OBE}
    \end{multline}
For a Lorentzian spectral density and identical emitter coupling ($g_{ik}=g_{k}$), the shifts are  $J_{nm}^{\b{OBE}}=-\f{g_{SE}^{2}\tilde{\Delta}}{{\tilde{\Delta}}^{2}+\kappa^{2}}$, and the decay rates read $\Gamma_{nm}^{\b{OBE}}=\f{2g_{SE}^{2}\kappa}{{\tilde{\Delta}}^{2}+\kappa^{2}}$, where $\tilde{\Delta} = \eta-\omega_L $.

The comparison between the LME and OBE shows similar qualitative behavior to that observed in the single emitter case. As depicted in Fig. \ref{fig:infidelity_multiple_emitters} Panel (a), for a monochromatic drive and moderate driving strength $g_{SL}/\Gamma_{FGR}\lesssim 1$, the LME exhibits exceptionally good infidelity at short times ($t\leq 1/\Gamma_{FGR}$), which increases to $1-F\sim 10^{-3.5}$ until $t\sim 2/\Gamma_{FGR}$ and to $1-F\sim 10^{-2.7}$ for longer times.
In the strong driving regime ($g_{SL}/\Gamma_{FGR}= 3.6$) the intermediate time infidelity remains above $10^{-2.5}$, and increases to above $10^{-2}$ at long times. At intermediate and long times the OBE outperforms the LME, achieving an improvement of around an order of magnitude in the infidelity. 
Contrastly, for a Gaussian pulse, the LME remains accurate even for very strong driving, Fig. \ref{fig:infidelity_multiple_emitters} Panel (b).

Figure \ref{fig:infidelity_multiple_emitters} Panel (b) presents the infidelity for weakly driven and non-driven initially two excited emitters. In contrast to the single emitter case, when the initial state includes multiple excitations, even in the absence of coherent driving the LME deviates from the exact result. Hence the obtained infidelity  $<10^{-3.3}$ throughout the whole studied time duration  ($t\leq 4/\Gamma_{FGR}$) constitutes a surprisingly good agreement with the exact result. 

 In the non-Markovian regime ($\Gamma_{FGR}/\kappa \approx 0.6$, $g_{SE}/\kappa = 0.55$), the LME accurately captures the dynamics until $t\approx 1/\Gamma_{FGR}$, where the infidelity increases above $10^{-2.5}$, see Fig. \ref{fig:infidelity_multiple_emitters_non_markov} Appendix \ref{apsec:driven_open_non_lin}. At longer times the prediction of the OBE surpasses the LME ($t>3/\Gamma_{FGR}$), which leads to the conclusion that a combination of the LME at short times and OBE at longer times may reliably capture the emitter dynamic beyond the Markovian regime.

Overall, the analysis of a single and two emitters evolution motivates the following interpretation of the dynamics.
At short times ($t<1/\b{2\Gamma_{FGR}}$), the non-linear effects remain small due to a minuscule two photons emission amplitude. It is enhanced once there is a substantial probability for the emitters to emit and reabsorb multiple photons ($t\sim 1/\Gamma_{FGR}$). This process leads to a rapid increase in infidelity, which reaches a maximum at $\sim t=1/\Gamma_{FGR}$.  At longer times the inherent memory loss of the environment suppresses the probability to reabsorb a photon, and the infidelity of the LME improves up to $10^{-4}$, for the studied time-duration ($t\leq4/\Gamma_{FGR}$).
The typical trend of the accuracy for a weakly driven single emitter, as well as the non-driven multiple emitters, suggests that the major source of inaccuracies of the LME is the incomplete characterization of the reabsorption process.

\subsection{Substantial short-time non-Markovian corrections}

\begin{figure}
    \centering
\includegraphics[width=8cm]{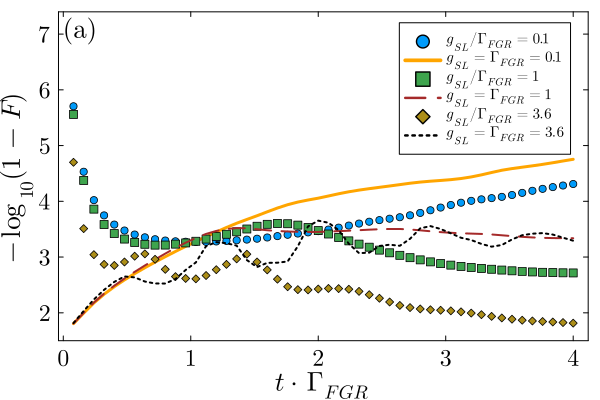}  
\includegraphics[width=8cm]{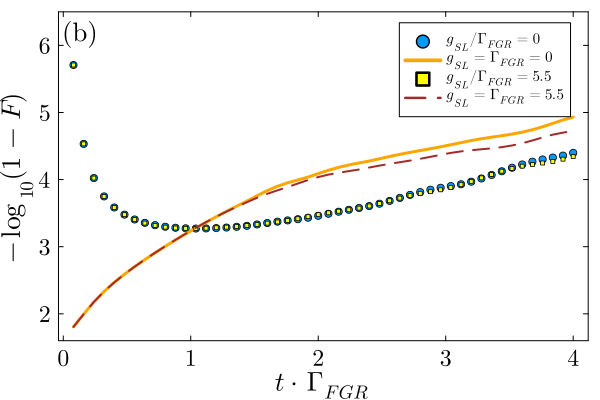}  \caption{Two emitter Markovian dynamics: (a) Monochromatic drive (b) Gaussian pulse and a non-driven system. Model parameters: $\ket{\psi_S\b 0} = \ket{e}_1\otimes\ket{e}_2$, $\Delta = 10^{-4}\omega_S$, $\Delta \omega_E = \kappa = 0.2 \omega_S$, $\Gamma_{FGR}=2\pi g_{SE}^2 D\b{\omega_S} = 6\cdot 10^{-2}\kappa$.}
\label{fig:infidelity_multiple_emitters}
\end{figure}

The excellent agreement of the LME with the exact result at short times ($t<1/2\Gamma_{FGR}$), and the large deviation from the OBE solution (Figs. \ref{fig:sigma_z_mono_non_Markov}, \ref{fig:infidelity_non_Markov}, \ref{fig:infidelity_multiple_emitters}) implies that even in the Markovian regime, for times $t<1/(2\Gamma_{FGR})$, the dynamics can be significantly influenced by the non-Markovian corrections. These non-Markovian effects manifest at times much longer than expected. In the studied Markovian models, the environment bandwidth is chosen as $\kappa = 0.2 \omega_S \approx 18 \Gamma_{FGR}$, hence one would expect that non-Markovian corrections occur at times comparable to $t\sim1/\kappa \sim 0.06 \Gamma_{FGR}$. Surprisingly, the results for single and two emitters suggest a contribution which influences the dynamics at much longer times. 
Remarkably, the non-Markovian effects are captured precisely by the short time behavior of the coefficients $\boldsymbol{f}_{NM}\b t$, ${\Gamma}^{\downarrow}_{ij}\b t$ (Eqs. \eqref{eq:f_NM} and \eqref{eq:dot_rho_s_lin}).

\section{Experimental realization}
\label{sec:experimental}


The major experimental challenge in the study of driven open system dynamics is the requirement for precise characterization, understanding and possible control of the environment. 
Namely, the environment in the experiment should precisely correspond to the theoretical model. This requires sufficiently slow dephasing of the environment and system due to unaccounted or uncharacterized noise. In practice, in order to observe most of the analysed phenomena it is sufficient to limit the dynamics to times up to $\sim 3/\Gamma_{FGR}$. Therefore, experimental platforms enabling loss rates of  $\gamma\sim 0.1 \Gamma_{FGR}$ are well suited for the task.  An additional desirable property is the ability to tune the system-environment coupling to the range of the environment's bandwidth $g_{SE}\sim \Delta\omega_E$, and maintain large Rabi-frequencies $g_{SL}\sim \Gamma_{FGR}$, while keeping the loss rates low.  
optical lattices \cite{de2008matter,navarrete2011simulating} and (iii) cavity QED \cite{kimble1998strong,walther2006cavity}. 

Patterning the structure of dielectric material enable control of the refractive index and engineering the dispersion relation of the confined photonic modes \cite{lopez2003materials,lodahl2015interfacing}. In this platform, a single system harmonic mode can be realized by introducing a point defect within a periodic dielectric material. The defect creates an effective highly tubable nano-cavity for the propagating light modes in the material. Alternatively, a 1D  (and 2D) tight-binding model environment,  studied in Sec. \ref{subsec:single harmonic}, can be realized by a series of defects forming a coupled resonator optical waveguide \cite{yariv1999coupled}. Couplings and system environment parameters are readily tuned by modifying the refractive index of the photonic crystal by mechanical means\cite{park2004mechanically}, thermooptic \cite{wang2011selective} and electrooptical \cite{du2004electrically} effects. 

Non-linear open systems can be readily realized by coupling solid-state, quantum dots, or natural atoms to a photonic crystal \cite{hung2013trapped,goban2014atom,lodahl2015interfacing}. A suitably designed  refractive index then enables mimicking the environment's spectral density function. Notably, to simulate the environment’s influence, it is sufficient to reproduce its spectral density or density of states within a limited frequency range on the order of the emitter linewidth, ${\sim\sb{\omega_S-n\Gamma_{FGR},\omega_S+n\Gamma_{FGR}}}$, where $n$ is a small integer. \cite{cohen1998atom}. This simplification allows the large tunability of the spectral properties of photonic crystals to be used to simulate the effects of band-less environments. Finally, for a 1D and 2D models the drive is realized by applying external lasers perpendicular to the  photonic crystal's surface \cite{lei2012quantum,englund2009quantum}. 

Coupling strengths of order $g_{SE}\sim 10 $ GHz, and a bandwidth of $J = \Delta \omega_E\sim 5 $ THz in the optical regime has been achieved \cite{douglas2015quantum}. As expected, inside the band the emitter decay is indeed Markovian ($g_{SE}/\Delta \omega_E\sim 10^{-3}$), nevertheless, non-Markovian dynamics emerge near the band edge (1D) or van Hove points (2D and 3D), where the group velocity vanishes and the density of states diverges \cite{gonzalez2017markovian,GonzalezTudela2018nonmarkovianquantum}. Rabi-frequency of the order of $g_{SL}\sim\Gamma_{FGR}$ have been obtained, allowing to saturate the transition \cite{zrenner2002coherent}.    The primary source of decoherence is the emission to free space, which can be low as $\gamma\sim 5 $ MHz \cite{douglas2015quantum},  reaching the desired dynamical regime $\Gamma_{FGR}\gg\gamma$ with $g_{SE}\sim \Delta \omega_E$.

In the cold atom realization, two metastable states of a natural atom interact with different optical lattice potentials. A thorough analysis provided in Refs. \cite{de2008matter,navarrete2011simulating,gonzalez2017markovian}, here we summarize the main features of the proposal. The primary system is realized by an atomic state which interacts with a strong laser, manifesting a deep lattice potential. This results in localized Wiener states at each potential well and inhibits tunneling to near by sites. The bosonic environment is realized by a second state, interacting with a shallow lattice potential which enables transition between neighboring sites. 
Importantly, the atomic density and magnetic fields are tailored to suppress any interaction between the atoms. Moreover, the atomic motion is confined to the lowest Bloch band in the collisional blockade regime, where at most a single atom occupies a potential site. The system environment coupling is controlled by laser driven direct optical or two-photon Raman transitions. Coupling strengths of the order of the tunneling rates can be achieved $g_{SE}\sim J = \Delta \omega_E \sim 10 $ KHz, while the strong driving regime can be reached using Rydberg atoms, manifesting Rabi-frequencies of the order of $g_{SL}\sim 10 $ MHz \cite{bloch2008many}. 

The dominant  sources of noise are scattering events induced by the trapping lasers and residual relaxation during the Raman transitions. These can be reduced to as low as $\gamma \sim 1$ Hz \cite{bloch2008many,schreiber2015observation,daley2008quantum}, reaching the desirable decoherence regime $\gamma/\Gamma_{FGR}\sim 10^{-3}$.



The well established field of cavity QED  
can also be 
employed to analyze the non-Markovian effects of the drive within the rotating wave approximation \cite{kimble1998strong,haroche2006exploring,walther2006cavity}. 
Following the analysis in Sec. \ref{subsec:pseudo}, leaky cavities enable a straightforward simulation of the driven open system dynamics.
The simplest case of emitters interacting with a Lorentzian spectral density environment, studied in Sec. \ref{sec:non_linear}, is realized by the standard QED setup, involving trapped emitters within a dissipative cavity. Coherent control is realized by directing lasers at the emitters. The strong coupling regime,  $g_{SE}\sim \kappa=\Delta\omega_E$, is achieved by using Rydberg atoms, confined in a high finesse cavity or near a whispering-gallery mode cavity,  while suppressing the emission to other electromagnetic modes ($\kappa$ is the cavity linewidth).

State-of-the-art optical cavities as the Fabry-Pérot  (whispering-gallery mode cavities) achieve linewidths of $\kappa\sim 1-10 $ kHz \cite{della2014extremely} ($100 \text{ kHz} – 1 $ MHz) with optimal coupling $g_{SE}\sim 100 $ MHz ($\sim $ GHz), Rabi frequencies up to $g_{SL}\sim 10 $ MHz for Rydberg atoms and free-space emission as low as $\gamma\sim 3$ MHz ($10$ MHz) \cite{kimble1998strong,buck2003optimal,foreman2015whispering}. For modest values of $g_{SE}=10\kappa= 100 $ MHz, one obtains $\gamma/\Gamma_{FGR}\sim 10^{-2}$, which allows capturing the non-Markovian effects of the drive. 

 Finally, circuit QED experimental setups  permit probing the non-Markovian effects of the drive and the interplay between coherent driving and dissipation beyond the rotating wave approximation \cite{blais2021circuit}. Namely, the so-called ultra-strong coupling regime can be achieved, where the coupling is engineered to be larger than the system frequencies \cite{bourassa2012josephson} ($g_{SE}/\kappa \sim 10^{2}-10^{4}$ with $\kappa$ as low as $\sim 10$ kHz and decay rates of order $\gamma\sim$kHz, and Rabi-frequencies up to $g_{SL}\sim $ MHz \cite{blais2021circuit}). Here, non-Markovian dynamics are realized by coupling a superconducting LC resonators (linear system) or superconducting qubits (non-linear system) to planar superconducting resonators, such as coplanar waveguides or 3D microwave cavities (the environment). An external drive can be readily applied using  microwave currents.


\section{Conclusions and perspective}


The interplay between coherent driving, dissipation and environmental memory effects generates a rich spectrum of dynamical phenomena. This study aimed to shed light on the mechanisms that govern the evolution of driven open quantum systems by comparing the performance of Markovian and non-Markovian master equations. Grasping the intricate interplay is vital not only for analyzing laser-induced phenomena in natural systems \cite{blankenship2011comparing,huelga2013vibrations,hubbard1964electron,holstein1940field}, but also for advancing technological applications \cite{verstraete2009quantum}.


Starting with a linear composite system, the exact drive-dependent non-Markovian effects were examined. Using the resolvent formalism, these corrections were directly linked to the spectral features of the environment. Notably, this effect is independent of the environment's temperature and arises from the initial state of the electromagnetic field (including the laser modes). This constitutes a distinct source of non-Markovianity, typically related to the spectral features of the environment.

Collective non-Markovian effects were subsequently studied, highlighting a novel cross-driving effect. Following, the general form of the exact master equation for a linear primary system, beyond the rotating wave approximation, was analyzed in Sec. \ref{sec:beyond_RWA}. The deduced master equation highlights that for arbitrary quadratic composite Hamiltonians and an initial Gaussian field state, the dissipation is independent of the drive.

An extension of the non-linear master equation, the linear master equation, was introduced (Sec. \ref{sec:lin_ME}) and compared to an exact pseudo-mode solution. The comparison yields several key conclusions:  
\begin{enumerate}[(i)]
    \item For an ensemble of emitters, non-Markovianity leads to modified time-dependence of the decay rates, bare system Hamiltonian, and external driving terms, which can be understood through a well-established analytical treatment of a corresponding linear system.
    \item  The non-Markovian corrections can manifest as cooperative dissipative and driving terms, as well as coherent interactions, in both linear and non-linear open systems.
    \item The excellent accuracy of the LME for driven single and two-emitter open system dynamics, moderate drives and non-Markovianity, suggests that the back-action of the environment on the system is only slightly affected by the non-linearity.
    \item Open system dynamics involving memory effects
    may be significantly affected by drive-dependent non-Markovian effects. This is manifested by a large magnitude of $\boldsymbol{f}_{NM}$ (Sec. \ref{sec:linear_sys}, Figs. \ref{fig:single_mode} and \ref{fig:cross_drive_effect}).
    \item  Even within the Markovian parameter regime, Eq. \eqref{eq:markov_cond}, when the bandwidths of the laser pulse and environment are comparable, non-Markovian effects lead to observable phase shifts (Sec. \ref{subsubsection:bandwidth}).
    \item   Non-Markovian effects can influence the relative accuracy of the evolution for times well beyond the typical decay time of environmental correlations.
    \item  For multiple non-driven emitters, the excellent accuracy of the LME provides a way to analytically investigate the multi-excitation regime. 
\end{enumerate}


The exceptional accuracy of the LME, at short times and within the Markovian regime (sections \ref{subsec:strong_driving_single_emiter} and \ref{subsec:two_emitters}), is especially relevant for information processing tasks including quantum error correction. A typical quantum information procedure involves applying control fields to a quantum device (the primary system), weakly coupled to its surrounding environment, between repeated measurements \cite{nielsen2010quantum}. To suppress errors, the measurements must be performed at time intervals which are small relative to the typical decoherence time ($\tau_{\text{meas}}< 1/\Gamma_{FGR}$). Crucially, after each set of measurements, the composite system-environment state collapses to a product state, before the inevitable emergence of new system-environment correlations. In this context, the analyzed short-time dynamics may correspond to the evolution between two subsequent measurements of a realistic quantum device. As a result, the accurate short-time description of the LME can be harnessed to engineer improved quantum gates, which account for the environmental noise \cite{kallush2022controlling}.

To illuminate the nature of interference between coherent driving and dissipation, we compared in Sec. \ref{sec:markov_regime} the performance of the Floquet, Adiabatic \cite{albash2012quantum}, and Time-Dependent master equations with the LME and its Markovian limit, the optical Bloch master equation. We found that the Markovian master equations incorporating the drive into the dissipation process (FME, AME, and TDME) deviate substantially from the exact solution. This comparison supports the conclusion that even for relatively strong driving, within the standard quantum optics regime, the dissipation remains independent of the drive.

The experimental considerations involved in realizing the drive-related open system phenomena were detailed in Sec. \ref{sec:experimental}. 
Analog simulation of open system dynamics proves to be a challenging task due to the inevitable interaction of the environment and primary system with uncontrolled sources of noise. Nevertheless, a number of state-of-the-art devices enable experimental confirmation of the proposed dynamical effects. 

An extension of the present analysis to finite-temperature thermal environments can be is achieved by using the general relation of the non-Markovian kinetic coefficients in Appendix \ref{apsec:ME_parameters}.
 The present analysis is however limited by the implicit application of the rotating wave approximation, leading to the considered Hamiltonian form. The analysis is therefore valid under the condition $g_{SL}\ll \omega_S$ \footnote{Interestingly, this condition coincides with the requirements that allow to the laser-matter interaction to be represented in terms of an explicitly time-dependent term in the Hamiltonian \cite{cohen1998atom}.}.

Finally, there are several possible extensions of the present analysis that are of significant interest. Future work may focus on including higher-order corrections of the non-linearity to the linear master equation and generalization to multi-level (beyond two-level) systems. Alternatively, the interplay between the counter rotating terms and coherent driving should be investigated  \cite{cao2021non}. Furthermore, the linear master equation can be applied to characterize the non-Markovian influence of environmental spectral features and laser pulses on collective phenomena such as superradiance \cite{dicke1954coherence,gross1982superradiance} and subradiance \cite{guerin2016subradiance}, the quantum state of emitted light \cite{parkins1995quantum,matsukevich2004quantum,porras2008collective,mitsch2014quantum,pizzi2023light,tziperman2024quantumstatelightcollective},  and dispersive interactions between quantum emitters \cite{buhmann2013dispersion,gonzalez2017markovian}.



\acknowledgements
I acknowledge insightful discussions with Ronnie Kosloff, Peter Salamon, James Nulton, Ohad Lib, Uri Peskin, Christiane Koch,  Alon Salhov, and Erikson Tjoa, as well as, Miguel Bello, Bennet Windt and Ignacio Cirac for stimulating discussions and for sharing their numerical codes with me. I am thankful for the support of the Quantum and Science Technology Program by the Council for Higher Education in Israel.

\section*{Appendix}

\appendix

\section{Clarification: linear and non-linear systems}
In the present context the term ``linear system'' refers to a bosonic (or fermionic) system with a quadratic composite (including both system and environment) Hamiltonian. As a consequence, the dynamics of such a composite system can be expressed by a set of linear coupled differential equations (the Heisenberg equation), where the number of equations corresponds to the number of harmonic modes. Moreover, in the analysis of the reduced system dynamics, inclusion of a Lindbladian term composed of jump operators in the set $L\in\{a,a^\dagger,a^\dagger a\}$ maintains the form of the equations of motion. 

In comparison, for a non-linear system, such as a two-level emitter coupled to bosonic modes, the composite operator algebra is non-compact, leading to an infinite set of coupled linear differential equations even for a single environment mode \cite{bruus2004many}.

\section{Non-Markovian master equation of linear systems}
\subsection{Non Markovian kinetic coefficients}

\label{apsec:ME_parameters}
Considering a bosonic composite system, including a set of system modes coupled to a bosonic environment, initially in a thermal state with inverse temperature $\beta$. The decay rates and coherent parameters are the elements of the following matrices
\begin{gather}
    \boldsymbol{\Gamma}^{\downarrow}\b t=\dot{\boldsymbol{V}}-\dot{\boldsymbol{W}}\boldsymbol{W}^{-1}\b{
    \boldsymbol{V}+\boldsymbol{I}}-\b{\boldsymbol{V}+\boldsymbol{I}}\b{\dot{\boldsymbol{W}}\boldsymbol{W}^{-1}}^{\dagger} \label{apeq:A1}\\
    \boldsymbol{\Gamma}^{\uparrow}\b t
    =\dot{\boldsymbol{V}}-\dot{\boldsymbol{W}}\boldsymbol{W}^{-1}\boldsymbol{V}-\boldsymbol{V}\b{\dot{{\boldsymbol{W}}}\boldsymbol{W}^{-1}}^{\dagger} \nonumber\\
    \boldsymbol{\Omega}\b t=\f i2\sb{\dot{\boldsymbol{W}}\boldsymbol{W}^{-1}-\b{\dot{\boldsymbol{W}}\boldsymbol{W}^{-1}}^{\dagger}} \nonumber~~,
\end{gather}
with
\begin{equation}
\boldsymbol{V}\b{t} = \boldsymbol{T}\b t\boldsymbol{n}_{BE}\b{\boldsymbol{E}}\boldsymbol{T}^\dagger \b t~~
\label{apeq:V}
\end{equation}
\begin{equation}
    \boldsymbol{T}\b t = -i\int_0^t d\tau \boldsymbol{W}\b{t-\tau}\boldsymbol{R}e^{-i\boldsymbol{E}\tau }~~,
\end{equation}
where $\boldsymbol{n}_{BE}=1/\b{\exp\b{\beta \boldsymbol{E}}-1}$, ($\boldsymbol{E}=\text{diag}\b{\omega_1,\dots,\omega_{N_E}}$) is a diagonal matrix including the thermal populations, given by the Bose-Einstein distribution and $\boldsymbol{R}$ is the coupling matrix, determined by Eq. \eqref{eq:expample_H_cal}.

For a fermionic composite system the $\boldsymbol{\Gamma}^\downarrow$ is slightly modified 
\begin{gather}
    \boldsymbol{\Gamma}^{\downarrow}\b t=-\dot{\boldsymbol{V}}+\dot{\boldsymbol{W}}\boldsymbol{W}^{-1}\b{
    \boldsymbol{V}-\boldsymbol{I}}+\b{\boldsymbol{V}-\boldsymbol{I}}\b{\dot{\boldsymbol{W}}\boldsymbol{W}^{-1}}^{\dagger}~~.
\end{gather}
while the form of $\boldsymbol{\Gamma}^{\uparrow}$ and $\boldsymbol{\Omega}$ remains identical. In addition, $\boldsymbol{n}_{BE}\b{\boldsymbol{E}}$ in \eqref{apeq:V} is replaced by a Fermi-Dirac distribution.

\subsection{Markovian dynamics}
\label{apsec:markov_dyn}
The Markovian decay rates are obtained by first applying the Sokhotski-Plemelj theorem:  $\lim_{\eta\ra 0^+} \f{1}{x\pm i \eta} =\mp i\pi \delta\b x+{\cal P}\b{\f{1}{x}}$, and preforming the so-called single pole approximation, replacing the self-energy dependence on the frequency by its value at resonance $\boldsymbol{\Sigma}\b{E^+}\approx \boldsymbol{\Sigma}\b{\omega_S+i0^+}$ \cite{cohen1998atom}. The procedure leads to the following matrix elements 
\begin{gather}
    \sb{\boldsymbol{\Delta}_0}_{ij} = {\cal P}\sum_k \f{\eta_{ik}\eta_{jk}^*}{\omega_S-\omega_k}   \\
    \sb{\boldsymbol{\Gamma}_0}_{ij} = 2\pi\sum_k {\eta_{ik}\eta_{jk}^*} \delta \b{\omega_S -\omega_k} ~~,
\end{gather}
of the Hermitian coherent and non-coherent Markovian dynamical contributions $\boldsymbol{\Delta}_0$ and $\boldsymbol{\Gamma}_0$. Overall, the Markovian propagator is given by (Eq. \eqref{eq:W_markov})
\begin{equation}
    W\b t \approx \exp\sb{-i\b{\boldsymbol{M}+\boldsymbol{\Delta}_0}t-i\boldsymbol{\Gamma}_0t/2}~~.
\end{equation}

\subsection{Block matrix inversion formula}
\label{apsec:block_inversion}
For a square block matrix
\begin{equation}
\boldsymbol{M}=\sb{\begin{array}{cc}
\boldsymbol{A} & \boldsymbol{B}\\
\boldsymbol{C} & \boldsymbol{D}
\end{array}}    
\end{equation}
and invertable matrices, $\boldsymbol{A}$ and $\boldsymbol{Q}=\boldsymbol{A}-\boldsymbol{B}\boldsymbol{D}^{-1}\boldsymbol{C}$, the inverse matrix is given by 
\begin{equation}
\boldsymbol{M}^{-1}=\sb{\begin{array}{cc}
\boldsymbol{Q}^{-1} & -\boldsymbol{Q}^{-1}\boldsymbol{B}\boldsymbol{D}^{-1}\\
-\boldsymbol{D}^{-1}\boldsymbol{C}\boldsymbol{Q}^{-1} & \boldsymbol{D}^{-1}+\boldsymbol{D}^{-1}\boldsymbol{C}\boldsymbol{Q}^{-1}\boldsymbol{B}\boldsymbol{D}^{-1}
\end{array}}    ~~.
\end{equation}

\section{Beyond the rotating wave approximation}
\label{apsec:beyond_RWA}

The following section briefly describes the construction in Ref. \cite{ferialdi2016exact} and provides a detailed analysis of the generalization of the non-Markovian master equation for a driven linear system coupled to a Gaussian bosonic environment by a bi-linear coupling.  Einstein summation notation is applied throughout this section.

The reduced dynamics of a system for such a composite system is governed by a completely positive Gaussian superoperator ${\cal M}_t$, satisfying 
$$\rho_S\b t = {\cal M}_t\rho_S\b 0~~.$$
Diosi and Ferialdi deduced the most general form of the superoperator
$${\cal M}_t = {\cal T}\exp\b{-\int_0^t d\tau S^j_\Delta\b{\tau}\int_0^\tau ds B_j \b{\tau,s}}~~,$$
with 
$$ B_j\b{\tau,s} = D_{jk}^{\text{Re}}\b{\tau,s}S_{\Delta}^k\b s +2 i D_{jk}^{\text{Im}}\b{\tau,s}S_c^k\b{s}~~,
$$ 
using the notation
\begin{equation}
S_\Delta^j \equiv S_L^j-S_R^j~~~~~,~~~~~S_c^j \equiv \b{S_L^j+S_R^j}/2~~, 
\label{apeq:833}
\end{equation}
where $S_{L,R}^j$ act on $\rho_S$ from the right and from the left, respectively.  
Here $D^{\text{Re}}$ and $D^{\text{Im}}$ are the real symmetric and imaginary antisymmetric parts of the environment's correlation function
$$D_{jk} = \text{tr}_E\b{E_j\b{\tau}E_k\b{s}\rho_E\b 0}~~.$$
$S^j\b{\tau}$ and $E^j\b{\tau}$ are the system and environment interaction operators in the interaction picture,  (equivalent to the Heisenberg picture relative to the free dynamics).

In order to derive the associated dynamical generator ${\cal L}_t \equiv \dot{\cal M}_t {\cal M}_t^{-1}$, satisfying 
$$\dot{\rho}_S\b t = {\cal L}_t \rho_S\b t~~,$$
Ferialdi \cite{ferialdi2016exact} employs Wick's theorem and the Cauchy identity for the  product of two series to express  $\dot{\cal M}_t$ in terms of ${\cal M}_t$. 
The derivation leads to the dynamical generator (master equation) of the form
\begin{multline}
    \dot{\rho}_S\b t = -S_{\Delta}^j\b t\{\int_0^t ds_1 {\cal A}_{jk}\b{t,s_1}S^k_{\Delta}\b{s_1}\\ + 2i {\cal B}_{jk}\b{t,s_1}S_c^k\b{s_1}\} \rho_S\b t~~,
    \label{apeq:840}
\end{multline}
where 
\begin{equation}
{\cal A}_{jk}\b{t,s_1} = D_{jk}^{\text{Re}}\b{t,s_1}+\sum_{n=1}^\infty \alpha_{jk}^n\b{t,s_1}
\label{apeq:cal A}
\end{equation}
\begin{equation}
{\cal B}_{jk}\b{t,s_1} =  D_{jk}^{\text{Im}}\b{t,s_1}+\sum_{n=1}^\infty \beta_{jk}^n\b{t,s_1}   
\label{apeq:cal B}
\end{equation}
with 
\begin{multline}
    \alpha_{jk}^n \b{t,s_1}\b{-1}^n \{\int_{s_1}^{t} ds_2 \int_0^t dt_2 b_j^{n,l} \b{t_2,s_2}D_{lk}^{\text{Re}}\b{s_2,s_1}\\+\int_0^{s_1}ds_2 \int_0^t dt_2 a_{jk}^n \b{t_2,s_2}
    \label{apeq:alpha}
\end{multline}
\begin{multline}
    \beta_{jk}^n\b{t,s_1}\b{-1}^n \{\int_{s_1}^{t}ds_2 \int_0^t dt_2 b_j^{n,l}\b{t_2,s_2}D_{lk}^{\text{Im}}\b{s_2,s_1}~~,
    \label{apeq:beta}
\end{multline}
and
\begin{equation}
    a^{1}_{j,j_2}\b{s_1,s_2} =\wick{\c1 B_j\b{t,s_1}\c1 B_{j_2}\b{t_2,s_2}}~~,
     \label{apeq:a_1}
\end{equation}
\begin{multline}
    a^{n}_{j,j_2}\b{s_1,s_2} = \int_0^t dt_{n+1}\int_0^{t_{n+1}} ds_{n+1}\{a_{jk}^1\b{s_1,t_{n+1}}\\\times a_{kj_2}^{n-1}\b{s_{n+1},s_2}+a_{jk}^{n-1}\b{s_1,s_{n+1}}\wick{\c1 S_{\Delta}^k\b{t_{n+1}} \c1 B_{j_2}\b{t_2,s_2}\}}
    \label{apeq:a_n}
\end{multline}
\begin{equation}
    b_j^{1,j_2}\b{s_2,t_2} = \wick{\c1 B_j\b{t,s_1}\c1 S_{\Delta}^{j_2}\b{t_2}}
     \label{apeq:b_1}
\end{equation}
\begin{multline}
    b_j^{n,j_2}\b{s_1,t_2} =\int_0^{t}dt_{n+1}\int_0^{t_{n+1}}ds_{n+1}b_j^{1,k}\b{s_1,t_{n+1}}\\ \times b_k^{n-1,j_2}\b{s_{n+1},t_2}
     \label{apeq:b_n}
\end{multline}
for $n\geq 2$.
The contraction is defined as
\begin{equation}
    \wick{\c1 B_{j_1}\b{t,s_1}\c1 S^{j_m}}_\Delta\b{t_m} = \sb{S_{\Delta}^{j_m}\b{t_m},B_{j_1}\b{t,s_1}}\theta\b{t_m-s_1}~~,
    \label{apeq:contraction}
\end{equation}
where $\theta$ is the unit-step function.

Importantly, the difference between the driven and non-driven system interaction terms, Eqs. \eqref{eq:320} and \eqref{eq:334} respectively, is only the scalar term $\bar{C}^j$. Hence, the contraction (Eq. \eqref{apeq:contraction}) does not depend on the drive. In turn, $a^1_{j,j_2}$ and $b^{1,j_2}_j\b{s_2,t_2}$ (Eqs. \eqref{apeq:a_1}, \eqref{apeq:b_1}) do not depend on the drive. Equations \eqref{apeq:a_n}, \eqref{apeq:b_n}, \eqref{apeq:alpha} and \eqref{apeq:beta} then imply that ${\cal A}_{jk}$ and ${\cal B}_{jk}$ are also independent of the drive. 

Returning to the derivation of the master equation, utilizing Eq. \eqref{eq:334} we next express $S_{\Delta}\b{s_1}$
and $S_{c}\b{s_1}$ in terms of the interaction operators at time $t$ 
\begin{equation}
    S_{\Delta}^{k}\b{s_{1}}={\cal C}_{q}^{k}\b{t-s_{1}}S_{\Delta}^{q}\b t+\tilde{{\cal C}}_{q}^{k}\b{t-s_{1}}\dot{S}_{\Delta}^{q}\b t
    \label{eq:B13}
\end{equation}
\begin{equation}
    S_{c}^{k}\b{s_{1}}={\cal C}_{q}^{k}\b{t-s_{1}}S_{c}^{q}\b t+\tilde{{\cal C}}_{q}^{k}\b{t-s_{1}}\dot{S}_{c}^{q}\b t+\bar{{\cal C}}^{k}\b t~~.
    \label{eq:B14}
\end{equation}
Substitution of Eqs. \eqref{eq:B13} and \eqref{eq:B14} into Eq. \eqref{apeq:840} leads to the interaction picture master equation 
\begin{multline}
    \dot{\rho}_{S}\b t=\bigg{(}\Gamma_{jk}\b tS_{\Delta}^{j}\b tS_{\Delta}^{k}\b t+\Theta_{jk}S_{\Delta}^{j}\b t\dot{S}_{\Delta}^{k}\b t\\-i\Xi_{jk}\b tS_{\Delta}^{j}\b tS_{c}^{q}\b t-i\Upsilon_{jk}\b tS_{\Delta}^{j}\b t\dot{S}_{c}^{k}\b t\\-i\Phi_{j}\b tS_{\Delta}^{j}\b t\bigg{)}\rho_S\b t~~,
\end{multline}
where
\begin{equation}
    \Gamma_{jk}\b t = -\int_0^t ds_1 {\cal A}_{jl}\b{t,s_1}{\cal C}_k^l\b{t-s_1}
\end{equation}
\begin{equation}
    \Theta_{jk}\b t = -\int_0^t ds_1 {\cal A}_{jl}\b{t,s_1}\tilde{\cal C}_k^l\b{t-s_1}
\end{equation}
\begin{equation}
    \Sigma_{jk}\b t = 2\int_0^t ds_1 {\cal B}_{jl}\b{t,s_1}{\cal C}_k^l\b{t-s_1}
\end{equation}
\begin{equation}
    \Upsilon_{jk}\b t = 2\int_0^t ds_1 {\cal B}_{jl}\b{t,s_1}\tilde{\cal C}_k^l\b{t-s_1}~~.
\end{equation}
\begin{equation}
    \Phi_{j}\b t = 2\int_0^t ds_1 {\cal B}_{jl}\b{t,s_1}\bar{\cal C}^l\b{t-s_1}~~.
\end{equation}
Finally, utilizing relation \eqref{apeq:833} and transforming back to the Schr\"odinger picture leads to the main result of Sec. \ref{sec:beyond_RWA} (Eq. \eqref{eq:nonRWA_ME})
\begin{multline}
        \dot{\rho}_S\b t= -i\sb{H_{S0}+H_d\b t,\rho_S\b t} -i\Phi_j\b t\sb{S^j,\rho_S\b t}\\ +\Gamma_{jk}\b t \sb{S^j,\sb{S^k,\bullet}}+\Theta_{jk}\b t\sb{S^j\sb{\dot{S}^k,\bullet}}
   \\ - i\Sigma_{jk}\b t\sb{S^j,\{S^k,\bullet\}}-i\Upsilon_{jk}\b t\sb{S^j\{\dot{S}^k,\bullet\}}~~.
\end{multline}



\section{Example of open system dynamics of a linear system}
\subsection{Detailed dynamical solution for a single modes coupled to a 1D photonic crystal}
\label{apsec:single_emit_1D}
The present appendix briefly describes the solution for the single emitter non-equilibrium Green function $W\b t$ Eq. \eqref{eq:W_int_single_emit}. The procedure closely follows the analysis given in Ref. \cite{gonzalez2017markovian}, see also Part III of Ref. \cite{cohen1998atom}  for an introductionary overview.
We work in th interaction picture relative to the bath's central frequency $\omega_E$. One begins by continuing the integral to the complex plane
\begin{equation}
    -\f{1}{2\pi i}\int_{C}\f{e^{-iz t}}{{z-{\omega_S}-{\Sigma}_{1D}\b{z}}}d z~,
    \label{apeq:W_int_single_emit}
\end{equation}
where the single emitter self energy $\Sigma_{1D}$ is given in Eq. \eqref{eq:Sigma_1D}.
The integrand is characterized by a continuous spectrum for $z\in\sb{-2J,2J}$ and two real poles in the band gaps: one at $\text{Re}\b{z}<-2J$  and the other at $\text{Re}\b{z}>2J$. 
The standard contour, including an arc in the bottom part of the complex plane cannot be used since the square root in $\Sigma_{1D}$ exhibits branch points at $z=\pm 2J$. These are circumvented by introducing two branch cuts from $\sb{\pm 2J,0}$ to $\sb{\pm 2J,\infty}$ correspondingly, and a contour which by pass these branch cuts. Moreover, the integrand exhibits a discontinuous jump when crossing the real axis in the regime $\text{Re}\b{z}\in \sb{-2J,2J}$. This is cured by an analytical continuation to the second Riemann surface in the transition to this region. The analytical continuation gives rise to the contribution of a complex pole to the integral.
Overall, three kinds of dynamical contributions are identified: Real poles, associated with matter-light bound states, which give rise to a coherent oscillation at the frequencies of the poles. The complex pole contributes an exponential decay, and the two branch cuts lead to a polynomial decay, scaling as $1/t^3$. 

The significance of each contribution depends on the coupling strength $g$ and detuning $\Delta = \omega_S-\omega_E$. For instance, the real poles will significantly influence the dynamics if they are near $\Delta$. While the branch cuts and complex poles significantly contribute to the dynamics near the band edge and within the band, correspondingly.

\subsection{Detailed dynamical solution for two modes coupled to a 1D photonic crystal}
\label{apsec:two_modes_exp}
In this section we present a solution for the propagator $\cal U$ associates with Hamiltonian \eqref{eq:H_example_2em}. 
The reflection symmetry enables decoupling $H$ into two commuting parts, each representing the dynamics of a collective system mode coupled to a distinct environment. The mathematical analysis is similar to the one presented in Refs. \cite{gonzalez2017markovian,windt2024fermionic}. 
We introduce the collective system and environment modes {$a_{\pm} =\b{a_{1}\pm a_{2}}/\sqrt 2 $} and $b_{k\pm}={\cal N}_{\pm}\sb{\b{e^{ikr}\pm e^{-ikr}}b_{k}+\b{e^{-ikr}\pm e^{ikr}}b_{{-k}}}$, with ${\cal N}_{\pm}=1/\b{2\sqrt{{1\pm\cos\b{2kr}}}}$, with $r=j_S a$, satisfying the canonical bosonic commutation relations. In terms of the collective modes the Hamiltonian separates to two commuting terms $H =\sum_{\alpha=\pm}H_\alpha$, with
\begin{equation}
 H_{\alpha}=\omega_{S}a_{\alpha}^{\dagger}a_{\alpha}+\sum_{k>0}\omega_{k}b_{k\alpha}^{\dagger}b_{k\alpha}+\sum_{k>0}g_{k\alpha}a_{\alpha}b_{k\alpha}^{\dagger}+\text{h.c} ~~,
\end{equation}
and $g_{k\alpha}=g\sqrt{2\b{1+\alpha\cos\b{2kj_S a}}/L}$.
As a result, the dynamical solution reduces to the calculation of the associated propagator elements
\begin{equation}
    a_\pm \b t= W_{\pm}\b t a_\pm \b 0 +\sum_{k}{T}_{\pm k}\b tb_{\pm k}~~. 
\end{equation}
The non-equilibrium Green functions of the collective modes, are calculated utilizing Eqs. \eqref{eq:U_int}, \eqref{eq:13}, \eqref{eq:14} and \eqref{eq:15}
\begin{equation}
  {W}_\pm\b t=
    -\f{1}{2\pi i}\int_{-\infty}^{\infty}{e^{-iE t}}G_{\pm}\b{E+i0^+}d E~.
    \label{apeq:W_pm}
\end{equation}
\begin{equation}
    T_{\pm k} =    \f{g_{k\pm}}{2\pi i}\int_{-\infty}^{\infty}\f{e^{-iE t}}{z-\omega_k}{G_{\pm}\b{E+i0^+}}d E
    ~~,
    \label{apeq:T_pm}
\end{equation}  
\begin{equation}
    G_{\pm} \b z=\f{1}{{z-\omega_S -{\Sigma}_\pm\b{z}}}~~,
\end{equation}    
and $\Sigma_{\pm}$ are the resolvents and self-energies of the collective modes.
By comparing Eq.  \eqref{eq:H_example_2em} with the general form Eq. \eqref{eq:H_comp1} and utilizing Eq. \eqref{eq:17} one obtains an explicit form for the self-energy 
\begin{equation}
 \Sigma_{\pm}\b{z,n_{12}}=\Sigma_{1D}\b z\pm\bar{\Sigma}_{12}\b{z,n_{12}}~~,
    \label{eq:Sigma}
\end{equation}
where $\Sigma_{1D}$ is the single mode self-energy given in Eq. \eqref{eq:Sigma_1D} and
\begin{multline}
    \bar{\Sigma}_{12}\b{z,n_{12}}=\f{g^{2}}{L}\sum_{k}\f{e^{ikn_{12}}}{z-\omega\b k}\\\ra\int_{-\pi}^{\pi}dk\f{e^{ikn_{12}}}{z+2J\cos\b{ka}}\\=\pm i \f{g}{\sqrt{4J^2-z^2}}\sb{\pm\f{z}{2J}\mp i \sqrt{1-\b{\f{z}{2J}}^2}}^{n_{12}}~~.
    \label{apeq:Sigma_bar}
\end{multline}
Here $\pm$ correspond to the case where $\text{Re}\b{z}\lessgtr 0$ and $n_{12}=2j_S a$ is the relative distance between the system modes \cite{gonzalez2017markovian} (see also \cite{windt2024fermionic} for an alternative form). The transition to the second line involves taking the continuum limit.

The integral of Eqs. \eqref{apeq:W_pm} and \eqref{apeq:T_pm} are evaluated by contour integration techniques described in \cite{gonzalez2017markovian,windt2024fermionic} and Appendix \ref{apsec:single_emit_1D}. 
In terms of the original modes we have 
\begin{equation}
    \boldsymbol{W}\b t= \boldsymbol{X}\text{diag}\b{W_+,W_-}\boldsymbol{X}
    \label{eqap:W_diag}
\end{equation}
\begin{equation}
   \boldsymbol{X}= \f 1{\sqrt{2}}\sb{\begin{array}{cc}
1 & 1\\
1 & -1
\end{array}} ~~.
\end{equation}  
The matrix $\boldsymbol{T}\b t$ is most conveniently defined in terms of its operation on the pair of Fourier operators $b_k,b_{-k}$; $$\boldsymbol{T}\b t:\{b_k,b_{-k}\}^T  \ra \b{\boldsymbol{X}\text{diag}\b{T_{+k},T_{-k}}\boldsymbol{X}} \{b_k,b_{-k}\}^T ~~.$$

Overall, the solution for $\boldsymbol{W}\b t$ determines the decay rates $\Gamma_{ij}^{\downarrow},\Gamma_{ij}^{\uparrow}$, dispersive shifts $\Omega_{ij}$ (Eq. \eqref{apeq:A1}) and driving term $k_i$ (Eq. \eqref{eq:k_t}) of the exact master equation Eq. \eqref{eq:dot_rho_s_lin}. 

Alternatively, the reduced dynamics can also be deduced from Eqs. \eqref{eq:solution} and \eqref{eq:13}. Introducing the operator valued vectors $\boldsymbol{a}=\{a_{-j_S},a_{j_S}\}^T$ and lattice momentum dependent vector of operators $\bar{\boldsymbol{b}}= \{b_{-\pi/a},\dots,b_{\pi/a}\}^T$, the Heisenberg picture dynamics of the system can be concisely expressed as 
\begin{multline}
    \boldsymbol{a}\b t= \boldsymbol{W}\b t \boldsymbol{a}\b 0 + \boldsymbol{T}\b t\bar{\boldsymbol{b}}\b 0 \\-i\int_0^t {\boldsymbol{W}\b{t-s}\boldsymbol{f}\b s} ds~~.
    \label{eq:solution2}
\end{multline}


\subsubsection{Non-Markovian driving term}
The form of the non-Markovian driving term $\boldsymbol{f}_{NM}$ can be written in a simple form. Utilizing Eq. \eqref{eqap:W_diag} one obtains
\begin{equation}
\boldsymbol{f}_{NM} \equiv \boldsymbol{X} \boldsymbol{\bar{f}}_{NM}  
\end{equation}
where $\boldsymbol{\bar{f}}_{NM} = \{\bar{f}_{NM+},\bar{f}_{NM-}\}^T$, and

\begin{multline}  
\bar{f}_{NM\pm}\b t=\int_0^t{W_\pm\b{t-\tau}\bar{{f}}_\pm\b{\tau}}\\-\dot{W}_\pm\b{W_{\pm}^{-1}}\int_0^t{W_{\pm}\b{t-\tau}\bar{f}_\pm\b{\tau}d\tau}~~.
\end{multline}

\section{Markovian Master equations}
\label{apsec:Markov_ME}

\subsection{Time-dependent master equation}
\label{apsec:TDME}
The following section concisely presents the time-dependent master equation (TDME), originally proposed in Ref. \cite{di2024time}. A combined approach, utilizing the projection operator technique along with a rescaling of the time by the system environment interaction leads a master equation of the form
\begin{widetext}
\begin{equation}
\f d{dt}\rho_{S}\b t=-i\sb{H_{S}\b t+H_{LS}\b t,\rho_{S}\b t}+\sum_{\alpha,\beta}\sum_{p}\gamma_{\alpha\beta,p}\b t\sb{L_{\beta,p}\b t\rho_{S}\b tL_{\alpha,p}^{\dagger}\b t-\f 12\{L_{\alpha,p}^{\dagger}\b tL_{\beta,p}\b t,\rho_{S}\b t\}} ~~,
\label{apeq:1372}
\end{equation}    
\end{widetext}
corresponding to a system-environment interaction term $H_I = \sum_{\alpha} A_\alpha \otimes B_\alpha$, with self-adjoint system and environment interaction operators $A_{\alpha}$ and $B_{\alpha}$. 
The Lindblad jump operators are given by 
\begin{equation}
L_{\alpha,p}\b t=U_{S}\b t\ket{m_{0}}\bra{m_{t}}A_{\alpha}\ket{n_{t}}\bra{n_{0}}U_{S}^{\dagger}\b t~~,
\label{apeq:1379}
\end{equation}
defined in terms of the instantaneous system eigenstates, $\{\ket{m_t}\}$ (of $H_S\b t$) and isolated system time-evolution operator, which satisfies
\begin{equation}
    i\dot{U_S}\b t = H_S\b t U_S\b t~~.
    \label{eq:Schro_U_S}
\end{equation}
The Lamb shift Hamiltonian reads
$$H_{LS}\b t=\sum_{\alpha,\beta}\sum_{p}S_{\alpha,\beta,p}\b tL_{\alpha,p}^{\dagger}\b tL_{\beta,p}\b t~~,$$
and the decay rates and Lamb shift frequency are given by the half-Fourier transform of the environmental correlation functions
\begin{equation}
\int_0^{\infty}{dx}R_{\alpha\beta}\b xe^{ix\b{E_{n}\b t-E_{m}\b t}}=\f 12\gamma_{\alpha,\beta,p}\b t+iS_{\alpha\beta p}\b t~~,
\label{apeq:int_R}
\end{equation}
$$R_{\alpha\beta}\b x=\text{tr}_{E}\b{e^{iH_{E}t}B_{\alpha}e^{-iH_{E}t}B_{\beta}\rho_{th}}~~,$$
where $p=\b{n,m}$ indexes the transition between the instantaneous eigenstates of $H_S\b t$, while  $\{E_{n}\b t\}$ are the corresponding instantaneous eigensvalues.  
The derivation required that the correlation function decay faster than $1/t^{a}$, with $a>0$. 

To simplify Eq. \eqref{apeq:1372} all the terms corresponding the same Linblidiand term are gathered together
\begin{multline}
    \f d{dt}\rho_{S}=-i\sb{H_{S}\b t+H_{LS}\b t,\rho_{S}}\\+\sum_{p}\Gamma_{p}\b t\sb{L_{p}\b t\rho_{S}\b tL_{p}^{\dagger}\b t-\f 12\{L_{p}^{\dagger}\b tL_{p}\b t,\rho_{S}\b t\}}~~,
    \label{apeq:TDME_simp}
\end{multline}
where
\begin{equation}
    \Gamma_{p}\b t=\sum_{\alpha\beta}\gamma_{\alpha\beta,p}\bra{m_{t}}A_{\beta}\ket{n_{t}}\bra{n_{t}}A_{\alpha}\ket{m_{t}}~~,
    \label{apeq:Gamma_p}
\end{equation}
and
\begin{equation}
    L_{p}\b t=U_{S}\b t\ket{m_{0}}\bra{n_{0}}U_{S}^{\dagger}\b t~~.
    \label{apeq:L_p}
\end{equation}

In the studied exemplary model (Sec. \ref{sec:markov_regime}) the isolated system dynamics, including the free dynamics and drive, are governed by the general Hamiltonian 
$$H_S\b t=\delta\b t\sigma_{z}+\Omega_{x}\b t\sigma_{x}+\Omega_{y}\b t\sigma_y~~,$$
with instantaneous eigen energies 
$E_{\pm}\b t=\pm\sqrt{\delta^{2}\b t+|\boldsymbol{\Omega}\b t|^{2}}~~,$
and corresponding eigenstates
\begin{gather}
  \ket{+_t}=\cos\b{\f{\theta\b t}2}\ket 1+e^{i\phi\b t}\sin\b{\f{\theta\b t}2}\ket 0 \label{apeq:1392}\\ 
    \ket{-_t} =-\sin\b{\f{\theta\b t}2}\ket 1+e^{i\phi\b t}\cos\b{\f{\theta\b t}2}\ket 0 \nonumber~~,
\end{gather}
where $\theta=\text{atan}\b{\f{|\boldsymbol{\Omega}|}{\delta}}$.
Here, $\boldsymbol{\Omega}=\{\Omega_{x},\Omega_{y}\}^{T}$, or alternatively $\boldsymbol{\Omega}=|\Omega|e^{i\phi}$.
The considered system-environment interaction (Eq. \eqref{eq:692}) can be expressed as 
\begin{equation}
    H_I =  \sum_k\b{g_k \sigma_+ b_k +g_k^*\sigma_- b_k^\dagger} = \sigma_x B_x +\sigma_y B_y~~.
\end{equation}
with 
\begin{gather}
 B_x = \f{1}{2}\b{B+B^\dagger }\nonumber\\   B_{y}=-\f i2\b{B^{\dagger}-B}~~.
 \label{apeq:B}
\end{gather}
Substituting Eq. \eqref{apeq:B}, \eqref{apeq:int_R},   \eqref{apeq:Gamma_p} and \eqref{apeq:L_p} into Eq. \eqref{apeq:TDME_simp} leads to 
\begin{equation}
\f d{dt}\rho_{S}\b t=-i\sb{H_{S}\b t+H_{LS}\b t,\rho_{S}\b t}+{\cal}D^{\b{TD}}\b t\sb{\rho_S\b t} ~~,   
\end{equation}    
where 
\begin{widetext}
    \begin{equation}
   {\cal}D^{\b{TD}}\b t\sb{\rho_S\b t} =  \sum_{k = 0,-}\Gamma_{k}\b t\sb{\Xi_{k}\b t\rho_{S}\b t\Xi_{k}^\dagger\b t-\f 12\{\Xi_{k}^\dagger\b t\Xi_{k}\b t,\rho_{S}\b t\}}~~,
\end{equation}
\end{widetext}
with
\begin{gather}
    \Xi_z \b t= U_S\b t \b{\ket{+_0}\bra{+_0}-\ket{-_0}\bra{-_0}}U_S^\dagger\b t \\
    \Xi_- \b t = U_S\b t \ket{-_0}\bra{+_0}U_S^\dagger\b t \nonumber ~~.\label{apeq:L_TD}
\end{gather}    
The decay rates are evaluated for  an environment at zero temperature, for which
\begin{multline}
\int_0^{\infty}{dx}R_{BB^\dagger}\b xe^{ix\omega_{nm}
\b t}=\\=\f{1}{2}\gamma_{nm}\b t +i S_{nm}\b t \\=    \pi \tilde{J}\b{\omega_{nm}\b t}+i{\cal P}\int_0^{\infty}{d\omega\f{\tilde{J}\b{\omega\b t}}{\omega_{nm}\b t-\omega\b t}}~~,
\label{apeq:1411}
\end{multline}
with $\omega_{nm}\b t=\b{E_{n}\b t-E_{m}\b t}/\hbar$.
Due to the original transformation to a rotating frame with respect to the laser carrier frequency, the spectrum and spectral density function of the environment  (due to a shift of the density of states) are shifted with respect to the original frame, $\Delta_k\in [-\omega_L,\infty)$, $\tilde{J}\b{\omega} = \b{g_{SL}^2\kappa/\pi}/\sb{\b{\omega-\b{\eta-\omega_L}}^2+\kappa^2}$.
Overall, we obtain the decay rates
\begin{gather}
  \Gamma_-\b t =\cos^4\b{\theta\b t/2} {2\pi} J\b{E_+\b t -E_-\b t} \nonumber\\
  \Gamma_z\b t =\f \pi4\sin^{2}\b{\theta\b t} J\b{0}~~.
  \label{apeq:Gamma_TDME}
\end{gather}
The Lamb shift term reads 
\begin{equation}
    H_{LS}\b t = S\b{E_+\b t -E_-\b t} \cos^4\b{{\theta\b t}/2}~~,
\end{equation}
where $S$ is evaluated by contour integration 
\begin{multline}
    S\b{\omega} ={\cal P}\int_{-\omega_L}^{\infty}{\f{\tilde{J}\b{\tilde{\omega}}}{\omega-\tilde{\omega}}d\tilde{\omega}}\approx {\cal P}\int_{-\infty}^{\infty}{\f{\tilde{J}\b{\tilde{\omega}}}{\omega-\tilde{\omega}}d\tilde{\omega}}\\ =\f{g_{SL}^{2}\b{\omega-\b{\eta-\omega_{L}}}}{\b{\omega -\b{\eta-\omega_L}}^{2}+\kappa^{2}}~~.
\end{multline}


\subsection{Adiabatic master equation}
The adiabatic master equation was religiously derived by Albash and Lidar \cite{albash2012quantum}, and can also be obtained by taking the adiabatic limit of the Time-dependent master equation, Appendix \ref{apsec:TDME}. In the adiabatic limit Eq. \eqref{eq:Schro_U_S} can be solved  by instantaneous diagonalization
\begin{equation}
    U_{S}^{\b{\text{adi}}}\b t = \sum_m e^{\int_0^t E_m\b{\tau}d\tau+\theta_m\b t} \ket{m_t}\bra{m_0}~~,
    \label{apeq:1453}
\end{equation}
where $\{\theta_m\b t\}$ are the Berry phases.
The decay remain the same as in the TDME (Eq. \eqref{apeq:Gamma_TDME}), while the Lindblad jump operators effectively become 
$$    L_{p}^{{\b{\text{adi}}}}\b t=\ket{m_{t}}\bra{n_{t}}~~.$$
In the studied model, this corresponds to a replacement 
\begin{gather}
 \Xi_z\ra \Xi_z^{\b{\text{adi}}} = \ket{+_t}\bra{+_t}-\ket{-_t}\bra{-_t}\\
\Xi_-\ra \Xi_-^{\b{\text{adi}}} = \ket{-_t}\bra{+_t} \nonumber~~. 
\end{gather}

\subsection{Floquet master equation}
The Floquet master equation for a two level, Eq. \eqref{eq:692}, has been studied extensively \cite{grifoni1998driven,szczygielski2013markovian,mori2023floquet}. Here, we focus and summarize the rigorous derivation presented Elouard et al. \cite{elouard2020thermodynamics}. The derivation starts by expressing the system interaction operators in the interaction picture in terms of Fourier components 
\begin{equation}
    \sigma_\pm \b t = \tilde{U}_S^\dagger\b t \sigma_\pm \tilde{U}_S\b t = e^{i\omega_L t}\sum_{\omega=0,\pm\Omega} \sigma_\pm \b{\omega}e^{i\omega t}~~,
\end{equation}
with $\sigma_{\pm}\b 0 = \f{g_{SL}}{\Omega}\Sigma_z$, $\sigma_\pm \b{-\Omega} = \mp \f{\Omega\pm \Delta}{2\Omega} \Sigma_-$ and $\sigma_{\pm}=\pm\f{\Omega\pm \Delta}{2\Omega}\Sigma_+$, where $\Sigma_z = \ket{+}\bra{+}-\ket{-}\bra{-}$, $\Sigma_- = \Sigma_+^\dagger = \ket{-}\bra{+}$, $\Omega=\sqrt{\Delta^2 +4 g_{SL}^2}$ and 
$\ket{\pm} = \pm\f{\sqrt{\Omega\pm\Delta}}{\sqrt{2\Omega}}\ket{e}+\f{\sqrt{\Omega\mp\Delta}}{\sqrt{2\Omega}}\ket g$ are the so-called ``dressed state'' which constitute eigenstates of the system Hamiltonian in the rotating frame
$$\tilde H_S \b t= \f{\Delta}{2} \sigma_z + g_{SL} \b{\sigma_+ +\sigma_-}~~.$$

The master equation in the rotating frame with respect to the driving frequency $\omega_L$ is of the form 
\begin{equation}
    \tilde{\rho}\b t = -i\sb{\tilde{H}_S,\tilde{\rho}\b t}+{\cal D}^{\b{FME}}\sb{\tilde{\rho}\b t}~~,
\end{equation}
where 
\begin{equation}
    {\cal D}^{\b{FME}} = {\cal D}_0 +{\cal D}_1 +{\cal D}_2
\end{equation}
with
\begin{gather}
    {\cal D}_0 = \b{\gamma_{0,\downarrow}+\gamma_{0,\uparrow}}{\cal D}_{\Sigma_z,\Sigma-z}\nonumber\\
    {\cal D}_1 = \gamma_{1,\downarrow}{\cal D}_{\Sigma_-,\Sigma_+} +\gamma_{1,\uparrow}{\cal D}_{\Sigma_+,\Sigma_-}\nonumber\\
    {\cal D}_2 = \gamma_{2,\downarrow}{\cal D}_{\Sigma_-,\Sigma_+}+\gamma_{2,\uparrow}{\cal D}_{\Sigma_+,\Sigma_-} \nonumber
\end{gather}
and kinetic coefficients
\begin{gather}
\gamma_{0,\downarrow}=\f{g_{SL}^2}{\Omega^2}\Gamma\b{\omega_L}\b{N\b{\omega_L}+1}\nonumber\\
\gamma_{0,\uparrow}=\f{g_{SL}^2}{\Omega^2}\Gamma\b{\omega_L}\b{N\b{\omega_L}}\nonumber\\
\gamma_{1,\downarrow}=\f{\b{\Omega+\Delta}^2}{4\Omega^2}\Gamma\b{\omega_L+\Omega}\b{N\b{\omega_L}+1}\nonumber\\
\gamma_{1,\uparrow}=\f{\b{\Omega+\Delta}^2}{4\Omega^2}\Gamma\b{\omega_L+\Omega}\b{N\b{\omega_L}} \nonumber\\
\gamma_{2,\downarrow}=\f{\b{\Omega-\Delta}^2}{4\Omega^2}\Gamma\b{\omega_L-\Omega}\b{N\b{\omega_L}+1} \nonumber\\
\gamma_{2,\uparrow}=\f{\b{\Omega-\Delta}^2}{4\Omega^2}\Gamma\b{\omega_L-\Omega}\b{N\b{\omega_L}}~~. \nonumber
\end{gather}

\section{Two emitter non-equilibrium Green function}
\label{apsec:two_emit}
In the following section we evaluate the non-equilibrium Green function (matrix) $\boldsymbol{W}\b t$ for two modes $a_1$ and $a_2$ coupled to a bosonic environment, characterized by a Lorentzian spectral density.
The Green function governs the dynamics of the modes. Hence, for an environment initially in the a vacuum state it satisfies
\begin{equation}
    \sb{\begin{array}{c}
\mean{a_{1}\b t}\\
\mean{a_{2}\b t}
\end{array}}=\boldsymbol{W}\b t\sb{\begin{array}{c}
\mean{a_{1}\b 0}\\
\mean{a_{2}\b 0}
\end{array}}~~.
\label{apeq:1702}
\end{equation}
This relation enables evaluating $\boldsymbol{W}$ by using the Heisenberg equations of motion of the pseudo-mode solution. Since the pseudo-mode (Sec. \ref{subsec:pseudo}) provides an accurate simulation of the system expectation values, we calculate the dynamics using the pseudo-mode solution of a non-driven system and deduce the Green function. 
In the pseudo-mode framework, the operator dynamics of the studied model are governed by  (Eq. \eqref{eq:9171}) 
\begin{equation}
    \f{dO}{dt} = i\sb{H_{SR},O}+2\kappa {\cal D}_{c,c^\dagger}^{\ddagger}\sb{O}
\end{equation}
with $H_{SR}\b t =\omega_S \sum_{i=1,2}a^\dagger_i a_i+\sum_{i=1,2}f_i\b ta^\dagger_i+f_i^*\b t a_i$.
Defining the rotating operators vector $\tilde{\boldsymbol{v}}\b t = e^{i\omega_L t}\{a_1\b t,a_2\b t,c\b t\}^T$ one obtains
$\dot{\tilde{\boldsymbol{v}}} = -i{\cal M}\tilde{\boldsymbol{v}}$, with
\begin{equation}
{\cal M}=\sb{\begin{array}{ccc}
\Delta_{1} & 0 & \lam\\
0 & \Delta_{2} & \lam\\
\lam & \lam & \eta'
\end{array}}
\end{equation}
and $\eta' = \eta-\omega_L-i\kappa$.
The solution for $\tilde{\boldsymbol{v}}$ is readily obtained in terms of the propagator $\exp\b{-i{\cal M} t}$
\begin{equation}
\tilde{\boldsymbol{W}}\b t=\sb{\exp\b{-i {\cal M} t}}_{1:2,1:2} = \sb{\begin{array}{cc}
\tilde{W}_{11} & \tilde{W}_{12}\\
\tilde{W}_{21} & \tilde{W}_{22}
\end{array}}
\end{equation}
with
\begin{multline}
    \tilde{W}_{11}\b t=\tilde{W}_{22}\b t=\f 12e^{-it\Delta}+2\lam^{2}\times\\\bigg{(}\f{e^{it\b{-\Delta-\eta'+\chi}/2}}{8\lam^{2}+\b{\Delta-\eta'}\b{\Delta-\eta'+\chi}}\\+\f{e^{-it\b{\Delta+\eta'+\chi}}}{8\lam^{2}-\b{\Delta-\eta'}\b{-\Delta+\eta'+\chi}} \bigg{)}
\end{multline}
\begin{multline}
    \tilde{W}_{12}\b t=\tilde{W}_{21}\b t=-\f 12e^{-it\Delta}+2\lam^{2}\times\\\bigg{(}\f{e^{it\b{-\Delta-\eta'+\chi}}}{8\lam^{2}+\b{\Delta-\eta'}\b{\Delta-\eta+\chi}}\\+\f{e^{-it\b{\Delta+\eta'+\chi}}}{8\lam^{2}-\b{\Delta-\eta'}\b{-\Delta+\eta'+\chi}}\bigg{)}~~,
\end{multline}
and $\chi=\sqrt{\b{\Delta-\eta'}^{2}+8\lam^{2}}$.
Finally, in the non-rotated frame, we have $\boldsymbol{W}\b t=e^{-i\omega_L t} \tilde{\boldsymbol{W}}\b t$.
\section{Additional figures of driven open non-linear systems}
\label{apsec:driven_open_non_lin}

    {\centering
    \includegraphics[width=8cm]{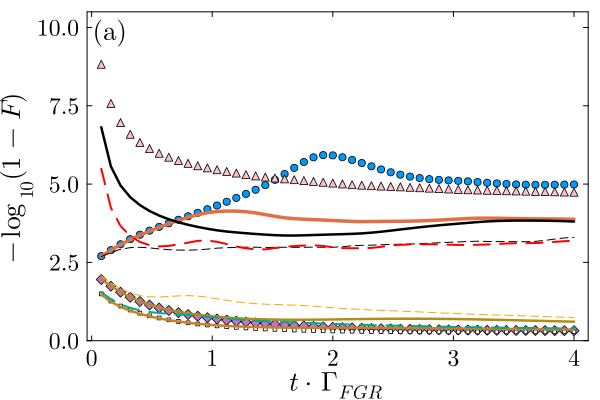}
    \includegraphics[width=8cm]{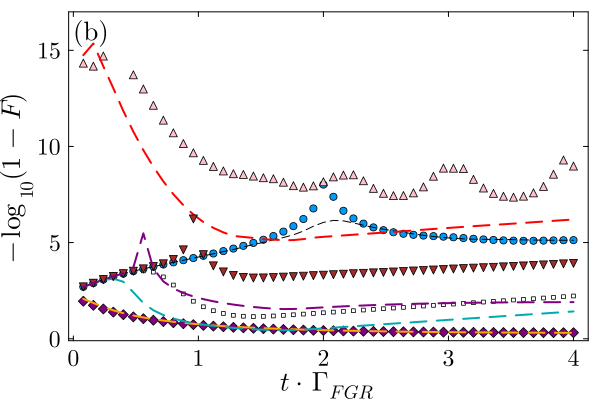}
    \captionof{figure}{Comparison of Markovian master equations: for a monochromatic laser (a) and Gaussian pulse (b).  Notation:  $g_{SL}= 0.1,1,3.6 \Gamma_{FGR}$ of LME correspond to pink up directed triangle markers, black continuous and red dashed lines; OBE: blue circle markers, red continuous and black dashed lines; FME: purple diamonds, light brown continuous and orange dashed lines. AME: small white square markers (appear on top of the purple diamonds), wide continuous orange and cyan dashed lines. For the monochromatic laser the TDME coincides with the AME, while for the Gaussian pulse the results for $g_{SL}=\Gamma_{FGR}$ of the LME and OBE overlap with the $g_{SL}=10^{-7}\omega_S$ markers. In Panel (b) the red down directed triangular markers and the purple dashed lines correspond to $g_{SL}=0.1 \Gamma_{FGR}$ and $g_{SL}=3.5 \Gamma_{FGR}$, respectively. Model parameters: $\Delta = 10^{-4}\omega_S$, $\Gamma_{FGR} \sim g_{SE}^2 D\b{\omega_S} \approx 1.7\cdot 10^{-3}\omega_S $, $\Delta \omega = \kappa = 0.2\omega_S$, $\eta = \omega_S$ and an initial composite state $\ket{\psi\b 0} = \sb{\b{\ket{g}+\ket{e}}/{\sqrt 2}} \otimes \ket{\text{vac}}$.}
\label{fig:infidelity_Markov}}

    \centering
\includegraphics[width=8cm]{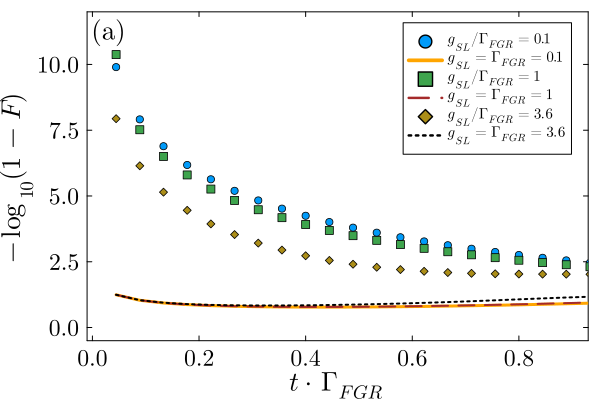}  
\includegraphics[width=8cm]{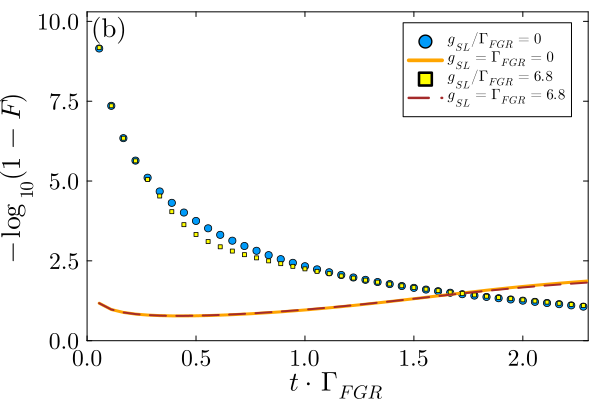} 
\captionof{figure}{Two emitter non-Markovian dynamics. Model parameters: $\ket{\psi_S\b 0} = \ket{e}_1\otimes\ket{e}_2$, $\Delta = 10^{-4}\omega_S$, $\Delta \omega_E = \kappa = 0.03 \omega_S$, $\Gamma_{FGR}=2\pi g_{SE}^2 D\b{\omega_S} = 0.55 \kappa$.}
\label{fig:infidelity_multiple_emitters_non_markov}

\bibliographystyle{plainnat}

\end{document}